\documentclass[%
 reprint,
superscriptaddress,
showpacs,
showkeys,
 amsmath,amssymb,
 prd,
]{revtex4-1}
\usepackage{graphicx}% Include figure files
\usepackage{dcolumn}% Align table columns on decimal point
\usepackage{bm}% bold math
\newcommand{\ssr}{Space Science Review}
\newcommand{\mnras}{MNRAS}
\newcommand{\aap}{Astronomy \& Astrophysics}
\newcommand{\jcap}{JCAP}

\begin{document}

\preprint{APS/123-QED}

\title{Dealing with missing data: \\An inpainting application to the MICROSCOPE space mission.}

\author{Joel Berg\'e}
 \email{joel.berge@onera.fr}
\affiliation{ONERA - The French Aerospace Lab, 29 avenue de la Division Leclerc, 92320 Ch\^atillon, France}

\author{Sandrine Pires}
\affiliation{
 Laboratoire AIM, CEA/DSM--CNRS %\\
 Universit\'e Paris Diderot %\\
  IRFU/SAp, CEA Saclay %\\
   Orme des Merisiers %\\
    91191 Gif-sur-Yvette, France
}%

 \author{Quentin Baghi}
 \affiliation{ONERA - The French Aerospace Lab, 29 avenue de la Division Leclerc, 92320 Ch\^atillon, France}
 
 \author{Pierre Touboul}
 \affiliation{ONERA - The French Aerospace Lab, 29 avenue de la Division Leclerc, 92320 Ch\^atillon, France}
  
  \author{Gilles M\'etris}
  \affiliation{Geoazur (UMR 7329), Observatoire de la C\^ote d'Azur Bt 4, 250 rue Albert Einstein, Les Lucioles 1, Sophia Antipolis, 06560 Valbonne, France}

\date{\today}% It is always \today, today,

\begin{abstract}
Missing data are a common problem in experimental and observational physics. They can be caused by various sources, either an instrument's saturation, or a contamination from an external event, or a data loss. In particular, they can have a disastrous effect when one is seeking to characterize a colored-noise-dominated signal in Fourier space, since they create a spectral leakage that can artificially increase the noise. It is therefore important to either take them into account or to correct for them prior to e.g. a Least-Square fit of the signal to be characterized. In this paper, we present an application of the {\it inpainting} algorithm to mock MICROSCOPE data; {\it inpainting} is based on a sparsity assumption, and has already been used in various astrophysical contexts; MICROSCOPE is a French Space Agency mission, whose launch is expected in 2016, that aims to test the Weak Equivalence Principle down to the $10^{-15}$ level. We then explore the {\it inpainting} dependence on the number of gaps and the total fraction of missing values.
We show that, in a worst-case scenario, after reconstructing missing values with {\it inpainting}, a Least-Square fit may allow us to significantly measure a $1.1\times10^{-15}$ Equivalence Principle violation signal, which is sufficiently close to the MICROSCOPE requirements to implement {\it inpainting} in the official MICROSCOPE data processing and analysis pipeline. 
Together with the previously published KARMA method, {\it inpainting} will then allow us to independently characterize and cross-check an Equivalence Principle violation signal detection down to the $10^{-15}$ level.

%\begin{description}
%\end{description}
\end{abstract}

%Use showkeys class option if keyword
\maketitle

%\tableofcontents

\section{Introduction} \label{intro}

Data acquisition in physics experiments is prone to intermittent cuts or errors. For instance, the observation strategy or commissioning periods, or maintenance works, may lead to gaps in the acquisition. Additionally, some events, either they are physical or instrumental, can lead to invalid data points that are {\it a posteriori} discarded, causing gaps in the data: examples of these kinds of gaps can be found in weak gravitational lensing, where cosmic rays and saturated pixels are masked (e.g. \cite{dejong15}), or in gravitational wave detection (e.g. \cite{vallisneri15}), where seismic events must be cut out.

Missing data complicate the data analysis by causing a spectral leakage in the frequency domain. This leakage is due to the convolution of the Fourier transform of the observational window with the Fourier transform of the complete data, causing the power from high-power regions to spread over the frequency domain.
It can significantly hamper data analysis, as it has been shown in fields as different as Cosmic Microwave Background (CMB) analyses (e.g. \cite{starck13, planck14, rassat14}), weak gravitational lensing (e.g. \cite{pires09, liu14, shirasaki14}), asteroseismology (e.g. \cite{garcia14, pires15}) or tests of fundamental physics (e.g. \cite{baghi15}). 
Optimizing the analysis by taking into account irregular data sampling is particularly important in the case where small signal-to-noise (S/N) deterministic signals are to be detected and characterized, as will be the case e.g. for the MICROSCOPE  (Micro-Satellite \`a tra\^in\'ee Compens\'ee pour l'Observation du Principe d'Equivalence, \cite{touboul01,touboul09, touboul12, berge15a}) or LISA Pathfinder \cite{mcnamara08,armano09} missions.
Efforts are underway to take missing data into account in data analysis, either by correcting them or designing data analysis methods insensitive to them (e.g. \cite{hikage11, vanderplas12} in weak lensing, \cite{perotto10} in CMB analyses, \cite{pires15} in asteroseismology, \cite{baghi15} for MICROSCOPE).
Among those efforts, {\it inpainting}, a method based on sparsity \cite{elad05} has been developed in the last ten years, with a special emphasis first on 2D cosmology \cite{pires09, starck13}, then on 1D astronomy \cite{garcia14, pires15}. Thanks to its versatility, it can in principle be applied to many data analysis cases. 

MICROSCOPE aims to test the Equivalence Principle in space with a $10^{-15}$ precision (at the $1\sigma$ confidence limit). The science data will consist of time series of differential accelerations measured by onboard inertial sensors. 
The data analysis challenge is to detect and characterize a possible very-small S/N Equivalence Principle Violation (EPV) signal. Despite its cutting-edge technology, it is expected that the data will be contaminated by uncontrollable events (such as impacts with micro-meteorites \cite{hardy13a, baghi15}). We will have to discards the measurements corresponding to those events, therefore causing gaps in the usable time series, which we will need to take into account. The MICROSCOPE noise being strongly colored, it will be particularly subject to spectral leakage due to possible missing data when analyzing the data in Fourier space. A major part of the data analysis will consist in taking those missing values into account.

In this paper, we apply an {\it inpainting} reconstruction to mock MICROSCOPE data. In particular, we show how we are able to recover the Equivalence Principle violation signal after correcting missing data with {\it inpainting}, with no {\it a priori} assumption on the noise. Indeed, estimating an unknown colored noise is particularly difficult \cite{vitale14}, especially when it is corrupted by missing data \cite{baghi15}. Therefore, we show that using a generic {\it inpainting} method, we are able to reach the MICROSCOPE's scientific goals. 

This paper is organized as follows. Sect. \ref{sect_miss} summarizes how missing data affect a general measurement and describes some advanced techniques developed to correctly deal with them; then Sect. \ref{sect_inpainting} briefly presents {\it inpainting}.  In Sect. \ref{sect_microscope}, we summarize the MICROSCOPE mission. We apply {\it inpainting} to MICROSCOPE mock data in Sect. \ref{sect_micinpainting}; in particular, we show the gain in precision that we obtain in the evaluation of a possible Equivalence Principle violation signal with a Least Square fit when gaps are filled with {\it inpainting}. Sect. \ref{sect_wingeom} explores how {\it inpainting} depends on the observational window. We conclude in \ref{sect_conclusion}.

Unless otherwise noted, all quoted errors are $1\sigma$ errors.

\section{Missing data} \label{sect_miss}

\subsection{Effect on the measurement: spectral leakage}

Let $X(t)$ be the ideal regularly sampled complete time series, $Y(t)$ the measured incomplete time series and $M(t)$ the binary mask (with $M(t)=1$ if we have information at data point $X(t)$, $M(t)=0$ otherwise), such that $Y(t)=M(t)X(t)$. 
In the Fourier domain the multiplication of the mask becomes a convolution: the Fourier transform of the signal $\tilde X(f)$ is convolved by the Fourier transform of the mask $\tilde M(f)$,
\begin{equation} \label{eq_conv} 
\tilde Y(f) = \tilde{M}(f) \ast \tilde X(f),
\end{equation}

At a given frequency, this convolution naturally produces a spectral leakage from this frequency to the frequencies around, the spectral leakage being characterized by the Fourier transform of the mask (the spectral window).  The latter depending on the geometry of the mask, the spectral leakage also naturally depends on the mask geometry. For instance, a single hole produces a cardinal sine spectral (sinc) window, whose beam's width depends on the hole's width. For regularly distributed gaps, the Fourier transform of the mask is a Dirac comb, hence it introduces spurious peaks in the PSD \cite{garcia14, pires15}. The case of a random mask is more complicated; as it is of interest for MICROSCOPE,  we summarize it in the next subsection. 

\subsection{Randomly distributed missing values and colored noise}

It can be shown that the spectral window for randomly distributed gaps of the same size, for a time series of length $L$, is asymptotically given by:
\begin{equation} \label{eq_spwin_main}
\tilde{M}(\omega)  =
\left\{
\begin{array}{l}
 \left| \delta(\omega) - \frac{f_g L}{\sqrt{2\pi}} \right|, \hspace{8mm} (\omega = 0) \\
 \frac{1}{2\sqrt{2}} \frac{f_gL}{\sqrt{N}} \left|{\rm sinc}\frac{f_gL\omega}{2N}\right|, \hspace{4mm} (\omega \neq 0),
\end{array}
\right.
\end{equation}
where $N$ is the number of gaps, $f_g$ the total fraction of missing values, $\omega$ the frequency, and $\delta(\omega)$ is the Dirac delta function.

Fig. \ref{fig_smoothed_mask} shows an example of spectral leakage due to randomly distributed gaps which could exist for MICROSCOPE (see Sect. \ref{sect_microscope} for more details). The blue line represents the PSD of the noise in presence of gaps, obtained from the black curve by convolution with the mask (Eq. \ref{eq_spwin_main}). The remaining lines are discussed in Sect. \ref{subsect_techniques}.

A short proof of Eq. (\ref{eq_spwin_main}) is given in the Appendix, together with a numerical example.
From this equation, we can see that the spectral window corresponding to randomly distributed gaps of the same size has a complex dependence on the number of gaps and on the total masked fraction. In particular, far from its main peak ($\omega \rightarrow \infty$), the level of the spectral window depends only on the number of gaps $N$, and is independent of the total masked fraction $f_g$. Moreover, the loss of power in the main peak depends only on the total masked fraction.
Numerical experiments show that this remains true when all gaps do not have the same size, in which case the sinc pattern disappears, but its envelope's level remains unchanged. Therefore, the spectral leakage far from its origin depends only on the number of gaps.

This is of particular importance for MICROSCOPE, since we will look for a deterministic signal, dominated by colored noise, far into the side lobes of the spectral window (as we describe in Sect. \ref{sect_microscope}, the frequency range of the signal we seek is between $10^{-4}$ Hz and $10^{-3}$ Hz, while the spectral leakage originates from frequencies around 1 Hz).
We will explore this behavior in Sect. \ref{sect_wingeom}.

\subsection{Techniques to deal with missing data} \label{subsect_techniques}

A large number of studies have been presented in the literature that discuss various advanced solutions to the problem of missing data in time series. We summarize some of them below.
 
A widely used technique to estimate a spectral density on irregularly sampled data is the Lomb-Scargle periodogram \cite{lomb76, scargle82}, which is based on a sine-wave least-square fitting. Assuming that a time series with missing data can be regarded as an irregularly sampled data set, we could view it as a way to correct for missing data. However, as mentioned in \cite{pires15}, this technique is subject to false detections, and we checked that it does not allow us to correct for missing data in a noise-dominated time series.

The CLEAN algorithm \cite{roberts87, foster95} is based on an iterative technique to detect peaks in the power spectrum, by detecting and removing them by descending amplitude in the Fourier domain, until the power spectrum is consistent with pure noise. However, any error on the estimated characteristics of removed peaks (amplitude, frequency, phase) introduces significant errors on the resulting final periodogram. As mentioned in \cite{pires15}, this is especially true when the spectral leakage is very large. It therefore cannot be used in our case of interest.

Recently, Baghi et al. \cite{baghi15} proposed the KARMA (Kalman-AR Model Analysis) method, a general linear regression method able to deal with incomplete data affected by unknown colored noise. It is based on an autoregressive (AR) fit of the noise that is used to whiten the data through a Kalman filtering process. By doing so the algorithm constructs a good approximation of the best linear unbiased estimator, conditionally to the AR model. The process is divided in three steps.  First the AR parameters are estimated with Burg's algorithm \cite{dewaele00} adapted to discontinous data. The purpose of this procedure is to approximate an arbitrary noise spectral density with an AR model of suitable order. The second step is to build a vector of uncorrelated entries using the outputs of a Kalman filter. This avoids storing and inverting large correlation matrices. In the third step a least-squares regression is performed on the whitened data, providing a result with a quasi-minimal variance. The process may be iterated a few times so that both the estimated noise spectrum and the estimated regression parameters reach convergence. 

Finally, a possibility to significantly decrease the spectral leakage consists in smoothing the window function, hence making it continuous: replace the abrupt changes from 1 to 0 (and vice-versa) around the $i$th hole (from times $t_i^{start}$ to $t_i^{end}$) by a cosine function parametrized by a parameter $\tau$, such as
\begin{equation} \label{eq_smooth}
M(t) = 
\left\{
\begin{array}{l}
1 \,\, {\rm if } \,\, t < t_i^{start}-\tau \,\, {\rm or} \,\, t > t_t^{end}+\tau \\
0 \,\, {\rm if } \,\, t_i^{start} \leqslant t \leqslant t_i^{end} \\
\frac{1}{2} \left[1 + \cos\left(\frac{\pi(t-t_i^{start}+\tau)}{\tau}\right)\right] \,\,{\rm if } \,\, t_i^{start}-\tau \leqslant t < t_i^{start} \\
\frac{1}{2} \left[1 + \cos\left(\frac{\pi(t-t_i^{end}-\tau)}{\tau}\right)\right] \,\, {\rm if } \,\, t_i^{end} < t \leqslant t_i^{end}+\tau \\
\end{array}
\right.
\end{equation}

\begin{figure} 
\includegraphics[width=0.45\textwidth]{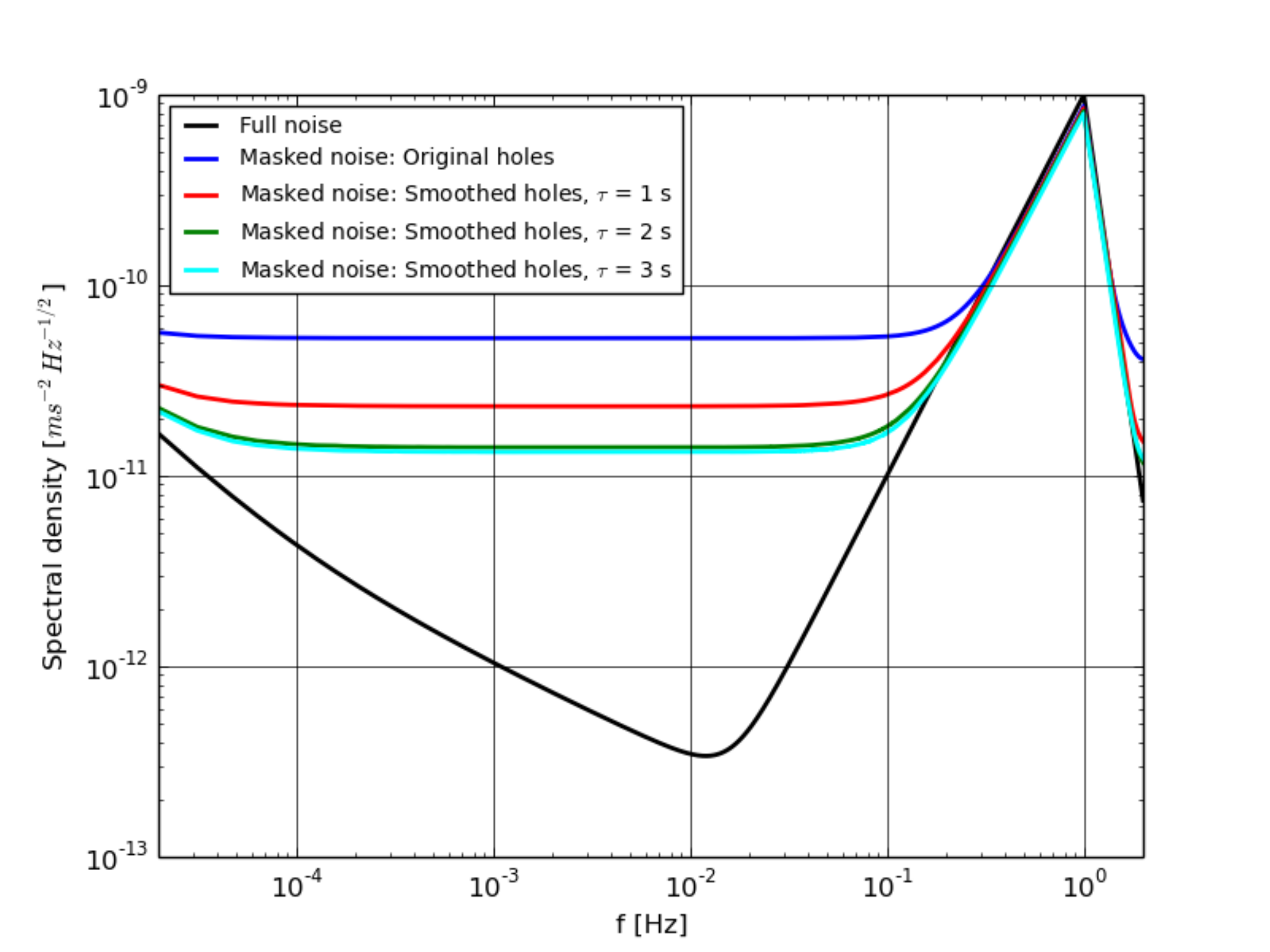}
\caption{Effect on the spectral leakage of smoothing the window function by defining it with Eq. (\ref{eq_smooth}). The blue line shows the spectral leakage with a binary mask, and other colored lines show the spectral leakage for smoothed windows, for varying $\tau$.}
\label{fig_smoothed_mask}       % Give a unique label
\end{figure}

As shown by Fig. \ref{fig_smoothed_mask}, re-defining the window function with Eq. (\ref{eq_smooth}) allows us to reduce the spectral leakage by up to a factor 5, depending on the value of $\tau$. However, when using this definition, the window is not binary anymore, thus preventing us from using {\it inpainting} (introduced in the next section): although reducing the spectral leakage, it cannot help reconstruct the missing information with {\it inpainting}. Moreover, Eq. (\ref{eq_smooth}) corresponds to some weighting of the data, that is less optimal for Least-Square estimation than KARMA. Although this definition may help recover a high S/N signal, it is insufficient for our purposes (as we show below, the deterministic signal we seek remains below the most reduced spectral leakage).

\section{Inpainting} \label{sect_inpainting}
In this paper, we investigate a different approach to deal with missing data which is called {\em inpainting} in analogy with the restoration process used in museums to restore deteriorated paintings. 
Inpainting techniques are well known in the signal processing field and are currently used to fill gaps by inferring a maximum information from the remaining data. 
The classical inpainting problem can be defined as follows:  having $Y=MX$, inpainting consists of recovering $X$ knowing $Y$ and $M$.

There is an infinite number of time series $X$ that can perfectly fit the observed time series $Y$. 
We propose to use an inpainting method, introduced by \cite{elad05}, that relies on a prior of sparsity, which can be easily applied to MICROSCOPE data.
This method makes use of the fact that all the representations of the signal are not equally interesting and pushes for sparse representations which makes information more concise and possibly more interpretable. 
This means that we have to find a representation $\alpha = \Phi^T X$ of the complete signal $X$ (of length $L$) in the dictionary $\Phi^T$ (of size $L \times L$) where most coefficients $\alpha_i$ are close to zero, while only a few have a significant absolute value.
For the MICROSCOPE application, $\Phi^T$ \footnote{The transpose notation is the common notation in the literature of the domain for the direct representation} can be the Discrete Cosine Transform (DCT) which is a decomposition into a set of oscillating functions like the Fourier transform. This DCT transform provides a sparse representation for the Equivalence Principle Violation signal and its calculation is efficient using Fast Cosine Transform algorithms that have the same computational complexity as Fast Fourier Transform algorithms.

Among all the possible solutions, we search for a unique solution by minimizing the number of large coefficient $\alpha$ in the selected dictionary $\Phi^T$ (i.e that provides a sparse representation of the complete data) while imposing that the solution is equal to the observed data within the intrinsic noise of the data.

Thus, the solution is obtained by solving:
\begin{equation}
\min  \| \alpha \|_1    \textrm{ subject to }  \parallel Y - MX   \parallel^2 \le \sigma^2,
%\min_\alpha{\parallel Y - M \Phi \alpha \parallel^2 + \lambda \parallel \alpha \parallel_1 },
\label{functional}
\end{equation}
where $||.||_1$ is the convex $l_1$ norm (i.e. $ || z ||_1 = \sum_k | z_k |$), $|| . ||$ is the classical $l_2$ norm (i.e. $|| z ||^2 =\sum_k (z_k)^2$) and $\sigma$ is the standard deviation of the noise in the observed time series.
The solution of such a regularization can be approximated through an iterative algorithm called Morphological Component Analysis (MCA) introduced by \cite{elad05}. This algorithm is based on a threshold that decreases exponentially at each iteration from a maximum value to zero. 
The conditions under which this algorithm provides an optimal and unique sparse solution to Eq. (\ref{functional}) have been explored by a number of authors (e.g. \cite{elad02,donoho03}). They showed that the proposed method is able to recover the sparsest solution provided this solution is indeed sparse enough in the representation $\Phi^T$ and the mask is sufficiently random in this representation.

The {\it inpainting} technique has already been used to deal with missing data in several astrophysics applications (e.g. \cite{pires09, jullo14, pires15}).

\section{The MICROSCOPE mission}  \label{sect_microscope}

Scheduled for launch in 2016, MICROSCOPE will test the Weak Equivalence Principle (WEP), a cornerstone of General Relativity, which states that two bodies in the same gravitational field experience the same acceleration, independently of their mass and composition. Thanks to a drag-free system and cutting-edge technology, MICROSCOPE will be able to test it down to the $10^{-15}$ level, 100 times better than the current best measurements \cite{will14}, by looking for a possible Equivalence Principle Violation modulated signal in several long time series of noisy data.

\subsection{Equivalence Principle Violation measurement principle}

The principle of the measurement is to compare the acceleration experienced by two free-falling test masses in the Earth's gravity field.
To this aim, MICROSCOPE embarks two ultrasensitive electrostatic differential accelerometers, each of these consisting of two coaxial cylindrical test masses whose motion is electrostatically constrained. The first accelerometer serves as a reference to demonstrate the experiment's accuracy: its test masses are made of the same material (Platinum-Rhodium PtRh10 alloy), so that it will not be subject to a true, physical EPV. The second accelerometer (`EP')  is used to test the WEP: its test-masses are made of different materials (PtRh10 and Titanium-Aluminium-Vanadium TA6V alloy). 
Any difference in the electrostatic servo-controlled acceleration applied to the masses to keep them relatively motionless is finely measured and provides the difference of the mass kinematic accelerations, a possible signal of the EP violation.

In orbit, the two masses of spherical inertia are precisely centered and they experience the same Earth gravitational field. Let us consider for simplicity a circular satellite orbit: in the instrument frame, the possible signal of violation, directed towards the Earth, is a sine whose frequency is defined by the orbital frequency, modulated by the spacecraft rotation.
%When the satellite is inertial or when it rotates about the normal axis to the orbital plane, the possible signal of violation, towards the Earth, is an orbital frequency sine in the instrument frame, modulated by the spacecraft rotation.

The measured signal in the MICROSCOPE experiment can thus be written as
\begin{equation}
y_{EP}(t) = \frac{1}{2} \mathcal{M}_{EP} \delta g_{EP}(t) + \mathcal{S}(t) + \mathcal{N}(t),
\end{equation}
where $\mathcal{M}_{EP}$ is an instrumental calibration factor \cite{touboul09}, $\delta$ is the EPV parameter we are aiming to detect and characterize, $g_{EP}$ is the Earth gravity field's projection on the measurement-axis, $\mathcal{S}$ represents systematics errors, and $\mathcal{N}$ is the statistical inertial sensor noise.

\begin{figure}
\includegraphics[width=0.45\textwidth]{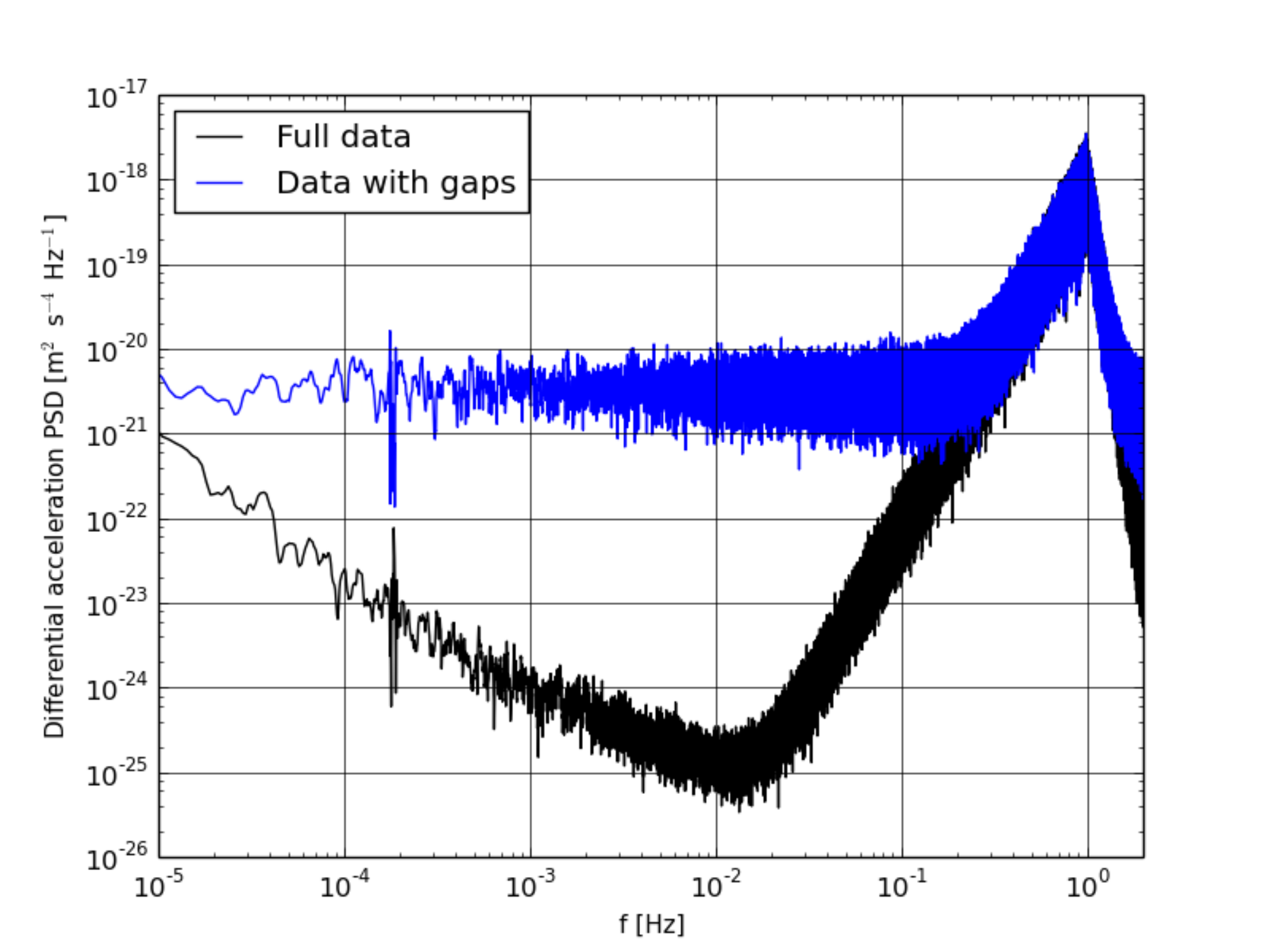}
\caption{Simulated MICROSCOPE's nominal noise PSD, smoothed with a 10-point-width Daniell filter (black). An example of a possible EPV is shown by the peak at $2\times10^{-4}$ Hz. The blue curve shows the PSD when gaps are present in the data.}
\label{fig_micnoise}       % Give a unique label
\end{figure}

The characteristics of the statistical noise $\mathcal{N}$ can be predicted by performance analyses \cite{touboul09}.
Its Power Spectral Density (PSD) can be decomposed in four regimes. A high-frequency increase as $f^4$ is due to the position detector noise; it is cut above 1 Hz by a Butterworth filter. The electronics noise is modeled as a $10^{-12}$ m/s$^2$/Hz$^{1/2}$ flat noise between $10^{-3}$Hz and $0.03$Hz. At lower frequency, the noise is dominated by the gold wire used to keep test-masses at a constant electric potential, with a PSD variation as $f^{-1}$. Lower-frequency thermal fluctuations are modeled as a $f^{-2}$ noise.
The noise is assumed to be Gaussian, and centered on 0.
On-ground performance measurements are in-line with those expectations.
The black curve in Fig. \ref{fig_micnoise} shows a simulation of the nominal noise as described above, smoothed with a 10-point wide Daniell filter. The peak around $2\times10^{-4}$ Hz is a possible $10^{-15}$ EPV (to prevent it from being erased by the smoothing, the frequency region surrounding it is smoothed by a 5-point wide Daniell filter).

Nuisance and calibration parameters $\mathcal{S}$ and $\mathcal{M}_{EP}$ can be either corrected for through careful modeling (e.g. gravity gradient tensor) or estimated by in-flight calibration (e.g. bias \cite{hardy13a, hardy13b}), or minimized by design of the instrument and of the satellite (e.g. inertial tensor, instrument's noise).

Thus, assuming perfect correction of systematics and instrument's calibration, we will measure a colored-noise ($\mathcal{N}(t)$) dominated sine-wave signal ($\delta g_{EP}(t)/2$). 

\subsection{Missing values and spectral leakage}

Different processes can cause data to be lost or not usable. A discussion about the sources of missing values in MICROSCOPE can be found in \cite{hardy13a, baghi15}. They can be divided in two main classes: (1) data transmission losses occur when a telemetry is not correctly received from the satellite; (2) saturated data occur when the instrument saturates (e.g. because of satellite's  multilayer insulation (MLI) coating crackles --as the temperature of the satellite's sides varies according to their orientation towards or opposite the Earth along the satellite's orbit--, or gas tank crackles --as the pressure of the tanks decreases as they are emptied by gas consumption--, or a micrometeorite impact). In the latter case, the data is not useable, and is therefore removed. 
In a worst case scenario, we expect that up to 3.5\% of the data may be masked: most expected holes come from tank crackles ($\approx$ 90\%), and MLI coating crackles ($\approx$ 10\%), and are shorter than a second; micrometeorite impacts are expected to be rare and create very short gaps; teletransmission losses are wider (up to several seconds) but very rare.

The combined effect of those masks on the simulated noise PSD (Eq. \ref{eq_conv}) is shown by the blue line in Fig. \ref{fig_micnoise}: there is an obvious spectral leakage from $f \approx 1$Hz to surrounding frequencies. As a result, the noise in the band $[10^{-4}-10^{-1}]$Hz is largely dominated by the spectral leakage from the high frequency noise: the deterministic signal we are looking for is therefore buried in the noise, while it would emerge clearly were all the data available.

\begin{figure}
\includegraphics[width=0.45\textwidth]{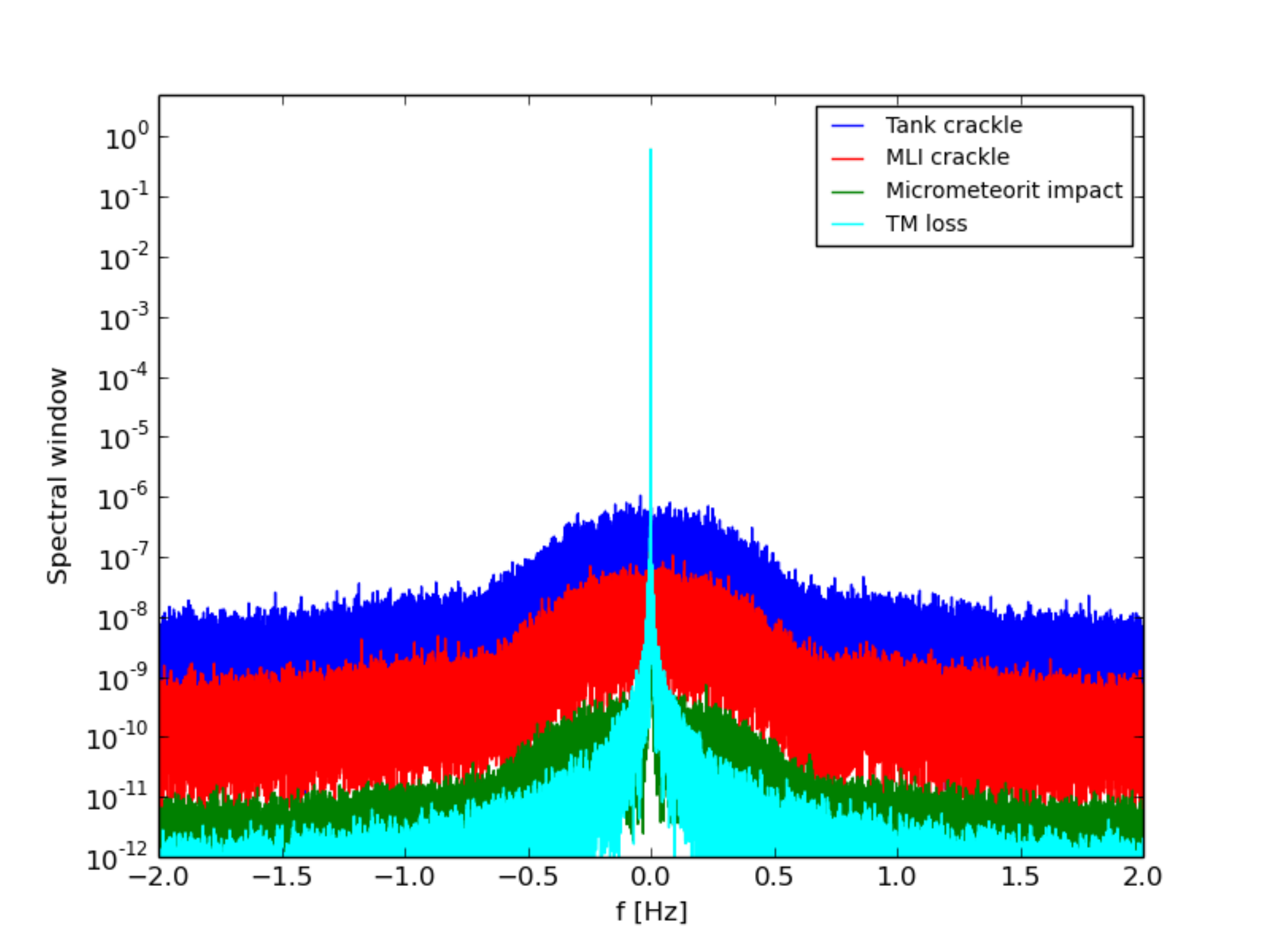}
\caption{Spectral windows corresponding to the possible gaps sources on MICROSCOPE.}
\label{fig_spwin}       % Give a unique label
\end{figure}

Fig. \ref{fig_spwin} shows the spectral window $\tilde{M}(f)$ for each type of gaps described above. Their convolution with the signal's PSD explains the leakage shown in Fig. \ref{fig_micnoise}. It is clear that the overall spectral window is dominated by that defined by tank crackles, while micrometeorite impacts and telemetry gaps have a negligible effect, their spectral window being two orders of magnitude smaller. 
As summarized in Sect. \ref{sect_miss}, far from the main peak, the spectral window level depends only on the number of gaps, and is independent of the total masked fraction, in the case of randomly distributed gaps (as for MICROSCOPE). This can be easily seen in Fig. \ref{fig_spwin}. Hence, the level of the leakage shown in Fig. \ref{fig_micnoise} depends only on the number of gaps (see also the Appendix).

\section{Inpainting application to MICROSCOPE} \label{sect_micinpainting}

\subsection{Simulations}

To assess {\it inpainting}'s performance on MICROSCOPE-like data, we design a suite of simulations with the assumption that all nuisance parameters are perfectly corrected for: hence, the signal consists of just a pure sine at a well known frequency and noise with PSD as shown by the black lines of Fig. \ref{fig_micnoise} and Fig. \ref{fig_mic_inpainted}: $y_{EP}(t) = \delta g_{EP}(t)/2 + \mathcal{N}(t)$.
For the sake of clarity in this section, and to have an acceptable S/N, we follow \cite{baghi15} and we set $\delta=3\times10^{-15}$.

We simulate typical ``inertial'' measurement sessions consisting of 120 orbits, where the satellite's attitude is kept fixed in an inertial frame. In this case, we expect an EPV signal at the orbital frequency, $f_{\rm orb}=1.8\times10^{-4}$ Hz, set by the mission design. Another type of measurement session consists in spinning the satellite about the axis normal to the orbital plane, so that the EPV signal is expected at a higher frequency (around $f_{EP} = 10^{-3}$ Hz): as the noise PSD is lower at this frequency (see Fig. \ref{fig_micnoise}), a 20-orbit integration is enough to detect a $10^{-15}$ EPV signal, were the data complete.

We define missing values in a worst case scenario, with 3\% of missing values, due to 260 tank crackles per orbit, 24 MLI coating crackles per orbit, 0.2 micrometeorite impacts per orbit and 0.05 telemetry loss per orbit. We set the mean duration of saturated data due to crackles and micrometeorite impacts to 0.75 seconds (corresponding to 3 data points), and the telemetry losses can vary from 1 second to 250 seconds. Gaps are not pre-defined, but their distribution is drawn randomly for each simulation, therefore we have access to their statistics only.

The black line in Fig. \ref{fig_mic_inpainted} shows the signal estimated PSD for a given simulation, when no mask is applied (i.e. the data is complete, regularly sampled), smoothed with a 20-point wide Daniell filter. Note that the smoothing decreases the apparent amplitude of the EPV peak.
The inset on the left shows a zoom around $f_{EP}$, with no smoothing.

\subsection{Inpainting}
\subsubsection{Missing data interpolation}

\begin{figure*}
\includegraphics[width=0.95\textwidth]{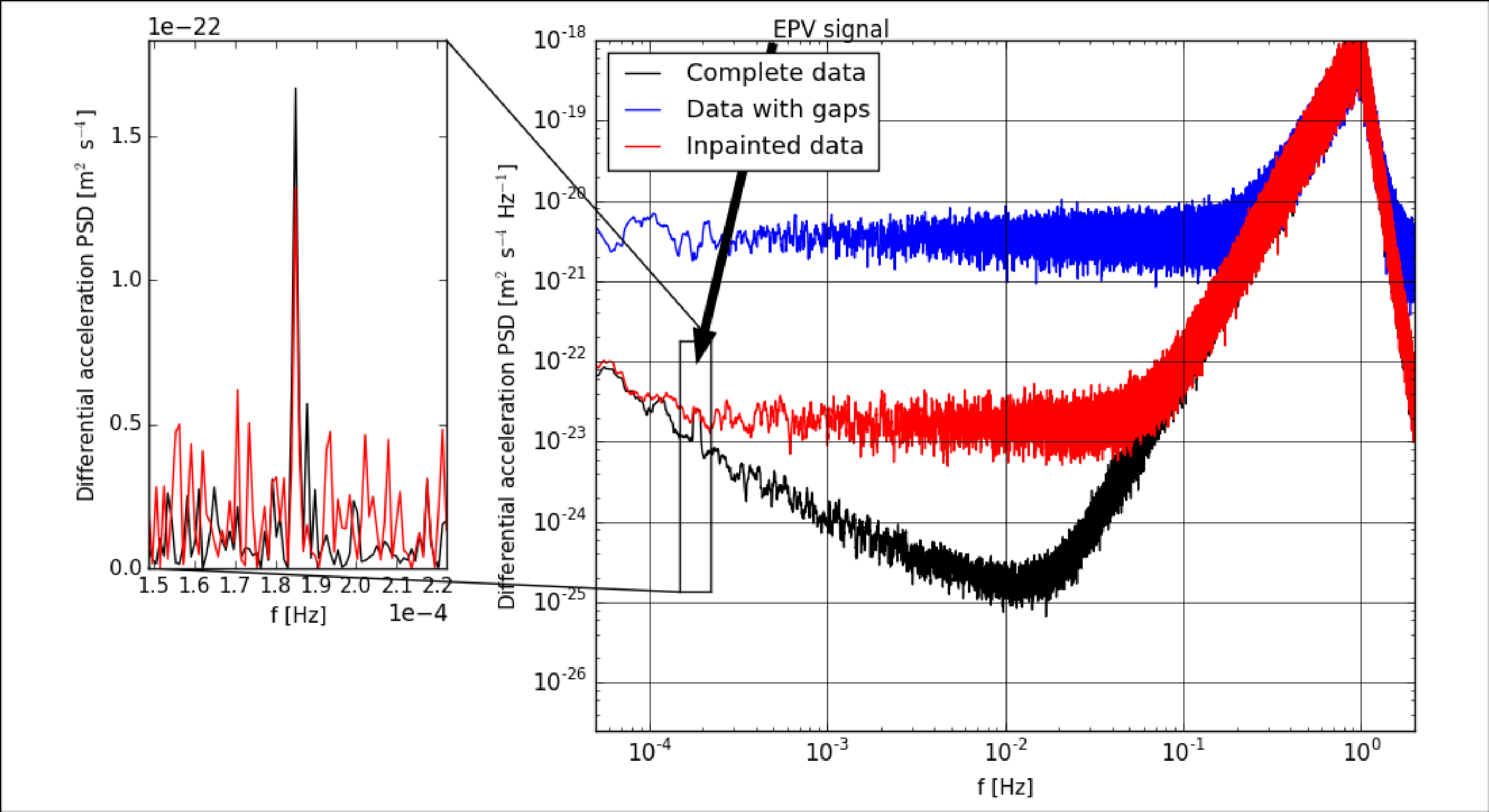}
\caption{MICROSCOPE differential acceleration PSD, smoothed with a 20-point Daniell filter, for a 120-orbit simulation. The black line shows the PSD when all the data is available, while the blue line shows the effect of missing values. The red curve shows the PSD after gaps are filled with inpainting. The inset on the left shows a zoom on the EPV signal, with no smoothing.}
\label{fig_mic_inpainted}       % Give a unique label
\end{figure*}

The blue curve in Fig. \ref{fig_mic_inpainted} shows the measured signal in the Fourier domain when some data is missing. 
We see the same spectral leakage as previously mentioned. It is clear that the EPV signal is completely swamped in the spectral leakage, thereby becoming extremely difficult to recover without any correction when working in Fourier space.

The red curve shows the PSD measured after reconstructing missing values with {\it inpainting}. 
It is noticeable that the original PSD is not recovered correctly at intermediate frequencies. This may be the sign that MICROSCOPE data brings {\it inpainting} to its limits: although the expected EPV signal is indeed sparse, the noise may not be sparse enough for the {\it inpainting} reconstruction to be optimal.
Despite this limitation, {\it inpainting} allows the EPV signal to emerge from the noise. It therefore becomes possible to detect and estimate the amplitude of this signal.

\subsubsection{EPV signal detection and characterization}

Once missing data have been reconstructed, we can use common techniques to detect and characterize the EPV signal: we indeed have a regularly sampled time series. 

For instance, we can characterize the EPV signal with a regression analysis. Since, in this work, we assume that the signal we are looking for is a pure sine wave, of known frequency and phase, we perform a simple Least-Square fit in the Fourier domain to the corrected data to estimate its amplitude. For more realistic signals, including some errors in the nuisance parameters, as well as perturbating signal (e.g. a low frequency drift), a more general regression technique will be needed, like a Maximum-Likelihood method or a MCMC technique, which will allow us to constrain more parameters than just the EPV signal.
Noise estimation is an upmost challenge in the Least-Square estimation of a signal (see e.g. \cite{vitale14, baghi15}). However, since in this work we are interested in evaluating the {\it inpainting}'s performance, we use a simple parametric fit of the noise, and do not look for the best possible noise estimator. We find that fitting the PSD of the noise as a sum of power laws (resembling the MICROSCOPE specifications) provides us with a good enough noise estimation. Nevertheless, we should emphasize that given this sub-optimal noise estimator, the errors on the Least-Square estimators we give below are to be considered conservative.

For the case of interest of Fig. \ref{fig_mic_inpainted}, where we simulated an EPV signal of $3\times10^{-15}$, we estimate $\delta=3.51\times10^{-15} \pm 1.36\times10^{-15}$ after {\it inpainting} correction. In the case with no gaps, our Least-Square fit brings $\delta_{\rm complete}=3.89\times10^{-15} \pm 0.90\times10^{-15}$, while we measure $\delta_{\rm gap}=22.0\times10^{-15} \pm 17.1\times10^{-15}$ before the {\it inpainting} correction. This is a single example which is not enough to characterize the performances of the method.

Therefore, we then create 300 similar simulations to allow us to perform a statistical analysis of our estimates. In the remainder of this section, the errors we quote are the rms of the Least-Square estimators estimated on our 300 simulations; in this way, we are able to quantify the combination of errors coming both from the {\it inpainting} interpolation and from the Least-Square estimation. 
We find that after correcting for the missing values with {\it inpainting}, on average, we measure an EPV of $\delta=2.69\times10^{-15} \pm 1.10\times10^{-15}$. With no {\it inpainting} correction, we would have measured $\delta_{\rm gap}=1.88\times10^{-15} \pm 13.5\times10^{-15}$ on average; in the case where no value is missing, $\delta_{\rm complete}=3.05\times10^{-15} \pm 0.76\times10^{-15}$. Therefore, {\it inpainting} allows us to have a significant (at better than $2\sigma$ confidence limit) measurement of a $3\times10^{-15}$ EPV signal, which would be impossible by simply performing an ordinary Least Square fit on the available data. Given our estimated $1\sigma$ statistical error, we can conclude that we may be able to characterize a possible EPV signal with a $1.1\times10^{-15}$ precision. Although slightly above MICROSCOPE requirements ($10^{-15}$ precision), this is good enough for us to decide to implement {\it inpainting} in the official MICROSCOPE data analysis pipeline.

\subsection{Comparison with KARMA} \label{sect_compkarma}

KARMA \cite{baghi15} and {\it inpainting} differ in their philosophy: {\it inpainting} fills gaps in a way that is independent of any physical model, thus allowing us to use an ordinary Least-Square method to estimate the EPV signal; KARMA, on the opposite, does not fill gaps but looks for an optimal estimator through a given model. 
Baghi et al \cite{baghi15} showed that the KARMA method applied to MICROSCOPE data allows us to measure a $3\times10^{-15}$ EPV signal at the 99\% confidence level in a similar worst case scenario than that used above. However, they assumed that the satellite was in ``spinning'' mode, where it spins about the axis normal to the orbital plane to increase the $f_{EP}$ frequency, therefore moving it to a region where the instrumental noise is lower than that in the ``inertial'' mode that we investigated above.

In order to better compare KARMA and {\it inpainting}, we applied KARMA to the 300 simulations used above to test {\it inpainting}. In that scenario, we find that KARMA allows  us to measure a $3\times10^{-15}$ EPV signal as $\delta_{\rm KARMA}= 3.06\times10^{-15} \pm 0.79\times10^{-15}$. Therefore, in the current ``inertial'' mode, KARMA's performances are similar to what was shown in \cite{baghi15}, and slightly better than those we estimated for {\it inpainting} above: on this mock data set, KARMA is more accurate and more precise than {\it inpainting}. However, as mentioned above, KARMA relies on an autoregressive model of the noise. This is a potential pitfall which might affect some data sets and which may be addressed in the future on additional data sets; in any case, it may require an independent data analysis to double check the results. {\it Inpainting}, although its performance is slightly poorer than KARMA's, with its different philosophy, is an ideal tool for this purposes. It has thus been decided that the MICROSCOPE pipeline will rely both on KARMA and {\it inpainting} to analyze data and confirm a possible EPV detection.

Moreover, {\it inpainting} is much faster than KARMA: the full data analysis (missing values reconstruction and Least-Square estimation) on a typical 120-orbit session takes a few minutes on a Mac Pro (3.5 Ghz 6-core Intel Xeon E5 processor, 64 GB RAM), while the full KARMA processing on the same machine and data takes about 45 minutes. This speed brought us to implement {\it inpainting} as the primary missing value reconstruction tool for daily instrument monitoring when MICROSCOPE flies; {\it inpainting} being less precise than KARMA is not important in this monitoring, where only large signals will be estimated, e.g. during calibration sessions \cite{hardy13b}.

\section{Inpainting dependence on the spectral window geometry: number of gaps and masked fraction} \label{sect_wingeom}

In this section, we probe the {\it inpainting} dependence on the mask geometry when running on MICROSCOPE mock data; that is, we investigate how it allows us to detect and characterize a low-frequency deterministic signal, affected by a loss of power and a spectral leakage from high-frequency noise due to missing values. 

To this aim, we design a set of simulations defining a grid in the ($N$, $f_g$) space, where $N$ is the number of gaps per orbit and $f_g$ the total fraction of missing values. As before, we simulate an $3\times10^{-15}$ EPV signal as measured by MICROSCOPE, under the assumption that all nuisance parameters are perfectly corrected for. 
We set the ($N$, $f_g$) couples as to explore the parameter space around the nominal MICROSCOPE values ($\approx 300$, $\approx$3\%). Although they are overly unlikely for MICROSCOPE, we also investigate extreme values ($f_g \geq $30\%) to increase the completeness of our exploration of {\it inpainting} performance.
For each couple ($N$, $f_g$), we create 100 simulations, run {\it inpainting} on them and measure the EPV signal with a Least-Square fit; our results are listed in Table \ref{tab_geoms}, where we report the mean measured EPV signal and statistical errors (multiplied by $10^{15}$), defined as before, as the rms of the Least-Square estimators for each set of 100 simulations. The couples with results noted ``N/A'' correspond to unphysical mask geometry (with a small masked fraction and a large number of gaps, which translates to gaps being smaller than one data point).

As mentioned in Sect. \ref{sect_miss} (see also the Appendix), the spectral leakage far from the frequencies where the spectral leakage originates (at those frequencies where we look for the EPV signal in MICROSCOPE) depends only on the number of holes, and is independent of the total proportion of masked data. We then expect {\it inpainting} to have an easier job recovering the noise level in the case where the number of gaps is small, and therefore we expect our Least-Square estimate to be more precise.
This is indeed what we observe in Table \ref{tab_geoms}: for a given total masked fraction $f_g$ the error on the estimated EPV signal increases with the number of gaps.
On the other hand, for a given number of gaps, the loss of power that affects the EPV signal itself because of missing values increases with the total masked fraction $f_g$ (Eq. \ref{eq_spwin_main}). Then, it will be more difficult for {\it inpainting} to reconstruct the EPV signal, and we expect the reconstruction to be biased. Indeed, we find that our Least-Square becomes less accurate as $f_g$ increases.

From this study, we can draw some rough upper limit on the allowed number of gaps and total masked fraction to significantly measure a given EPV signal with MICROSCOPE, when correcting for missing values with {\it inpainting}: for example, with 300 gaps per orbit, the total masked fraction should not exceed $\approx$ 3\%, which translates in gaps smaller than four data points (1 second) on average to characterize an EPV signal with a $1.08\times10^{-15}$ precision. For a smaller number of gaps, a significant fraction can be lost (up to more than 60\% with 30 gaps per orbit) while meeting the $10^{-15}$ MICROSCOPE requirement: this would imply large gaps (150 seconds) separated by large valid periods that provide uncontaminated information. Although such a gap configuration should not happen for MICROSCOPE, this number could be used to set the requirements of another, future mission.

\begin{table*}[t]%The best place to locate the table environment is directly after its first reference in text
\caption{\label{tab_geoms}%
EPV signal estimator and statistical error after {\it inpainting} correction, for different number of gaps $N$ and total masked fraction $f_g$. Those numbers correspond to the mean and the rms (multiplied by $10^{15}$) of the Least-Square estimators obtained on sets of 100 simulations for each couple ($N$, $f_g$).
}
\begin{ruledtabular}
\begin{tabular}{l|c|cc|cc|cc|cc|cc}
& \multicolumn{10}{c}{$N$ \textrm{ (per orbit)}} \\
\cline{3-12}
& & \multicolumn{2}{c}{30} & \multicolumn{2}{c}{100} & \multicolumn{2}{c}{300} & \multicolumn{2}{c}{1000} & \multicolumn{2}{c}{3000} \\
\cline{3-12}
& & $10^{15}<\hat\delta>$ & $10^{15}\sigma_\delta$ & $10^{15}<\hat\delta>$ & $10^{15}\sigma_\delta$ & $10^{15}<\hat\delta>$ & $10^{15}\sigma_\delta$ & $10^{15}<\hat\delta>$ & $10^{15}\sigma_\delta$ & $10^{15}<\hat\delta>$ & $10^{15}\sigma_\delta$\\
\colrule
& 1.5\% &  3.10 &  0.83 &  3.12  & 0.88 & 2.82 & 1.06 & N/A  & N/A  & N/A  & N/A\\
& 3\% & 2.96 & 0.83 & 2.79 & 0.86 & 2.89 & 1.08 & N/A  & N/A  & N/A  & N/A\\
$f_g$ & 15\% & 3.06 & 0.82 & 2.94 & 0.90 & 2.57  & 1.44 & 5.43 & 3.36 & 3.93  & 2.28 \\
& 30\% & 2.84 & 0.88 & 2.50 & 1.13 & 2.69  & 1.73 &  6.67 & 4.55 &  8.12 & 5.71 \\
& 60\% & 2.40 & 1.08 & 1.80 & 1.17 & 2.77 & 2.21 & 9.40  &  10.2 & 23.4 & 22.1 \\
\end{tabular}
\end{ruledtabular}
\end{table*}

\begin{figure} 
\includegraphics[width=0.45\textwidth]{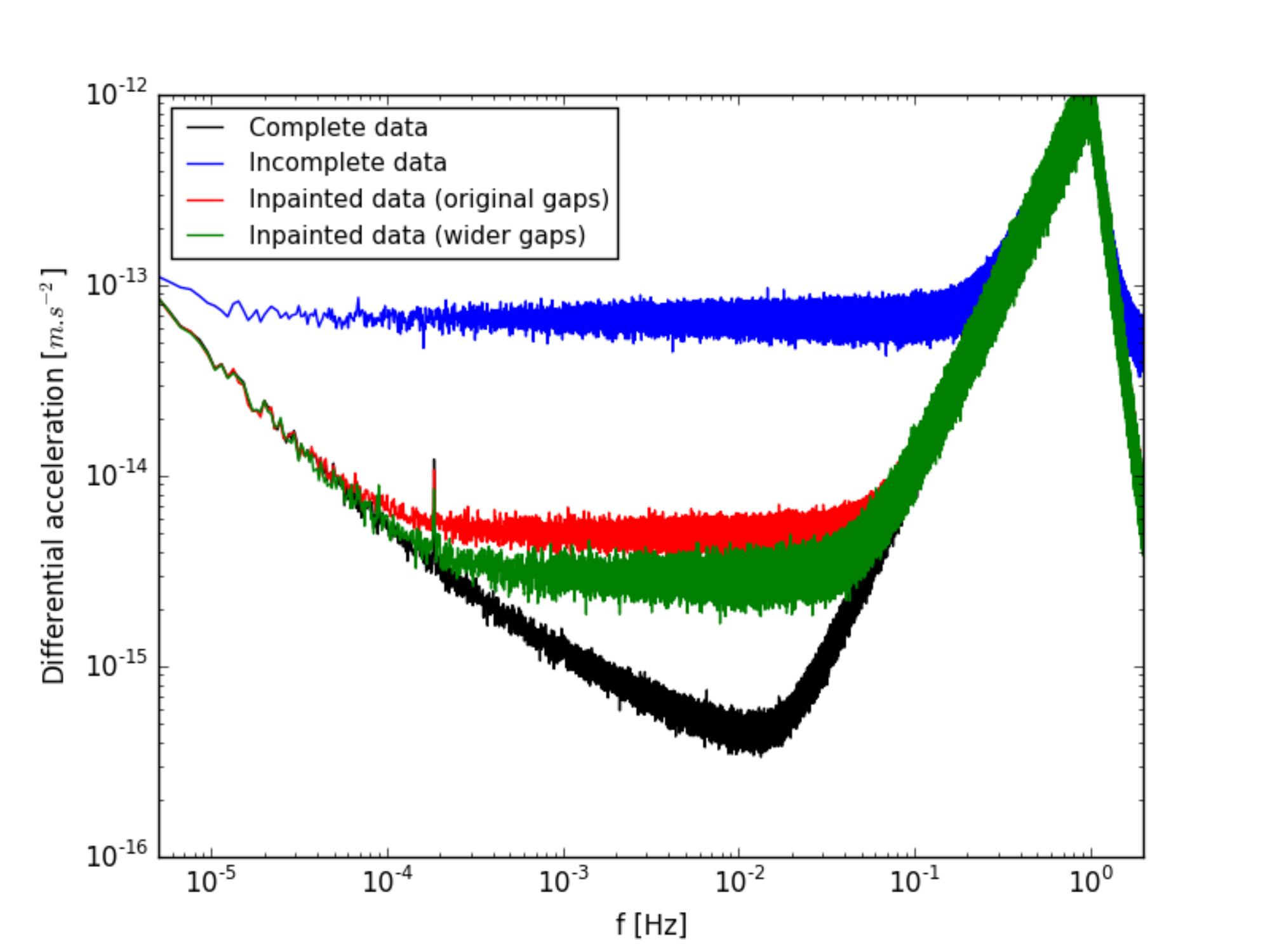}
\caption{Linear power spectrum of the mock MICROSCOPE differential acceleration, corrected with different masking strategies, averaged over 40 simulations. {\it Black}: complete data. {\it Blue}: data with missing values. {\it Red}: data corrected with {\it inpainting}, with no modification of the original gaps. {\it Green}: data corrected with {\it inpainting}, after gathering gaps closer than 30 seconds. Note the EPV peak at $1.8\times10^{-4}$ Hz.}
\label{fig_inpainted_widen}       % Give a unique label
\end{figure}

This study brings us to design the following strategy to improve the missing values reconstruction: gather gaps with their immediate neighbors if they are separated by less than a given (adhoc) distance, thus simplifying the spectral window's geometry at the expense of removing valid data. Doing so, the number of gaps decreases, although the total masked fraction increases. To illustrate this, we create a set of 40 simulations, identical to those used above. The average linear power spectrum of the differential acceleration for complete and gaped data are shown in black and blue in Fig. \ref{fig_inpainted_widen}, respectively. The red line shows the average linear power spectrum after the raw {\it inpainting} correction.
We then increase the size of gaps by gathering those that are closer than 30 seconds: on average, the number of holes is decreased by a factor of 5, while the masked fraction goes from 3\% to 50\%.
It is clear that as we decrease the number of gaps, the recovered noise after the {\it inpainting} correction decreases (green line in Fig. \ref{fig_inpainted_widen}), thereby improving the error on the Least-Square estimator; nevertheless, it is also clear from the figure that the amplitude of the recovered EPV signal peak decreases, thereby biasing the Least-Square estimator.
Therefore, removing valid data to improve the signal characterization, although it allows us to improve the precision of our estimator, degrades its accuracy: a trade-off should then be found, that optimizes the precision and the accuracy of the Least-Square estimate. Characterizing it is beyond the scope of this paper, but may be of interest to optimize the MICROSCOPE data analysis and will be done in a further work.

\section{Conclusion} \label{sect_conclusion}

We have presented an {\it inpainting} application to MICROSCOPE data to correct for missing values.
We first summarized the impact of missing data on a measured time series: in Fourier space, the observational window is convolved with the complete data, thereby creating a spectral leakage which artificially increases the noise at low frequency and hampers the detection and characterization of a small S/N deterministic signal. In particular, the use of an ordinary Least-Square estimation can be made impossible, as shown by \cite{baghi15}. Therefore, missing values need to be taken into account or corrected for in order to characterize a deterministic signal. This will be of upmost importance for the MICROSCOPE mission.

We then briefly mentioned some existing methods to deal with missing or irregularly sampled data; in particular, KARMA \cite{baghi15} has already been shown to perform well on MICROSCOPE data and reach the requirement to measure a $10^{-15}$ EPV signal with a $1\sigma$ confidence limit.
After shortly presenting {\it inpainting}, we applied it to mock MICROSCOPE data and showed that it allows us to correct for missing values in a worst-case scenario well enough to detect and characterize a $3\times10^{-15}$ Equivalence Principle Violation signal at better than the $2\sigma$ confidence level. Our estimated statistical $1\sigma$ error of $1.1\times10^{-15}$ is slightly bigger than required for the nominal MICROSCOPE performance ($10^{-15}$). 
This slight subperformance certainly comes from the fact that  although the expected EPV signal is indeed sparse in the DCT representation, the noise is not sparse enough, thereby limiting the {\it inpainting} reconstruction. The {\it inpainting} method reaches its limits with MICROSCOPE-type data. Further work to optimize {\it inpainting} in the MICROSCOPE case will be investigated in a future study. For instance, we could use a prior on the noise that could be inferred from a segment of the data without gaps. We are also currently investigating the feasibility of a hybrid technique between {\it inpainting} and KARMA, which is able to give an optimal estimator of the noise (Pires et al in prep). 

We then explored how {\it inpainting} depends on the number of gaps and on the total masked fraction. This allowed us to set requirements on those two parameters for {\it inpainting} to reach MICROSCOPE's requirements on the EPV measurement. On the one hand, we saw that the precision of the Least-Square estimation of the EPV signal after an {\it inpainting} reconstruction mostly depends on the number of gaps; on the other hand its accuracy decreases with increasing total masked fraction, the estimator becoming biased low.

Finally, we showed that {\it inpainting}'s performance is close to MICROSCOPE's requirements. As a consequence, we now have another tool to correct for missing values and characterize an EPV signal, that we can use in parallel with KARMA to reliably assert the detection and characterization of an Equivalence Principle violation signal.
This study brought us to include {\it inpainting} in the official MICROSCOPE's data processing and analysis pipeline. Since it is fast and model-independent, it is well adapted for a quick missing data correction, needed for day-to-day instrument's monitoring and calibration. It will also be used along-side KARMA to estimate the EPV signal, allowing us to cross-check our results with two independent techniques.

\begin{acknowledgements}
We thank Emilie Hardy, Manuel Rodrigues, Bruno Christophe and Alain Robert for useful discussions, as well as our anonymous referee for useful suggestions and comments.
This work makes use of technical data from the CNES-ESA-ONERA-CNRS-OCA Microscope mission, and has received financial support from ONERA and CNES.
We acknowledge the financial support of the UnivEarthS Labex program at Sorbonne Paris Cit\'e (ANR-10-LABX-0023 and ANR-11-IDEX-0005-02).
\end{acknowledgements}

\appendix* %remove the star if more than one appendix
\section{Spectral leakage, random gaps and colored noise}

In this Appendix, we prove Eq. (\ref{eq_spwin_main}) and give a numerical example. 

\subsection{Analytical proof} \label{sect_analytic}

We assume a time series of length $L \gg 1$, in arbitrary units. %Having $L \gg 1$, we can safely consider the time series to be infinite.
The window function that we aim to compute is defined in Sect. \ref{sect_miss}: $M(t)=1$ where data is available, $M(t)=0$ where data is missing.

We further assume that data is missing in $N$ non-overlapping holes, whose random distribution follows a uniform law. 
For simplicity, we assume that gaps are all of the same width $\Delta$.

The mask function can be written:
\begin{equation}
M(t)=1-\sum_{k=1}^N {\rm rect}_\Delta(t+n_kT),
\end{equation}
where $n_k$ is a uniformally distributed random variable such as $n_{k+1} - n_k > \Delta > 1$ (non-overlapping holes), so that $n_{k+1}/n_k>1$ and $T$ is a given (adhoc) time constant whose value is not important (since it is degenerate with the values of $n_k$).

Then, the spectral window is (with $\omega = 2\pi f$):
\begin{equation} \label{eq_tmp1}
\tilde{M}(\omega) = \left|\delta(\omega) - \frac{\Delta}{\sqrt{2\pi}} {\rm sinc}(\Delta\omega/2) \sum_{k=1}^N {\rm e}^{-i \omega n_k T}\right|.
\end{equation}

In the case $\omega=0$, the sum over $k$ in Eq. (\ref{eq_tmp1}) reduces to $N$ and the spectral window is:
\begin{equation} \label{eq_spwin_eq}
\tilde{M}(\omega)|_{\omega=0} = \left| \delta(\omega) - \frac{f_g L}{\sqrt{2\pi}} \right|,
\end{equation}
where we introduced the total masked fraction $f_g = N \Delta / L$.
For a given time series, it depends on the total masked fraction only: it corresponds to the loss of power due to missing values at a given frequency.

In the case $\omega \neq 0$,
\begin{equation} \label{eq_tmp}
\tilde{M}(\omega)|_{\omega \neq 0}= \frac{\Delta}{\sqrt{2\pi}} \left|{\rm sinc}\frac{\Delta\omega}{2}\right| \left|\sum_{k=1}^N {\rm e}^{-i \omega n_k T}\right|.
\end{equation}

To estimate the trigonometric sum with random frequencies $\sum_{k=1}^N {\rm e}^{-i n_k T \omega}$ for each $\omega$, we can define $\omega_k \equiv n_k T \omega$ and the complex variable $Z \equiv \sum_{k=1}^N {\rm e}^{-i \omega_k}$ (as mentioned above, the value of $T$ and $\omega$ do not impact the result, since $n_k$ is a random variable). Then, according to \cite{salem47}, and remembering that $n_{k+1}/n_k>1$ (so that $\omega_{k+1}/\omega_k>1$), the real and imaginary parts of $Z$ are approximately independent Gaussians with mean equal to zero and variance equal to $N/2$, for large $N$.
Therefore, its modulus $|Z|$ follows a Rayleigh distribution, with mean $<|Z|> = \sqrt{\frac{N}{2}}\sqrt{\frac{\pi}{2}}$ and variance ${\rm Var}(|Z|) = \frac{N}{2} \frac{4-\pi}{2}$.

Therefore, we can approximate, for every $\omega$, $\left|\sum_{k=1}^N {\rm e}^{-i n_k T \omega}\right| \approx \sqrt{\frac{\pi}{2}}\sqrt{\frac{N}{2}}$.

Then, the spectral window for irregularly spaced gaps on an infinite time series, is given by:
\begin{equation} \label{eq_spwin_neq}
\tilde{M}(\omega)|_{\omega \neq 0} = \sqrt{N} \frac{\Delta}{2\sqrt{2}} \left|{\rm sinc}\left(\frac{\Delta\omega}{2}\right)\right|.
\end{equation}

Returning to a finite time series of length $L \gg 1$, and substituting the total masked fraction $f_g$ to $\Delta$ in Eq. (\ref{eq_spwin_neq}), we finally obtain:
\begin{equation} \label{eq_spwinfg}
\tilde{M}(\omega)|_{\omega \neq 0}= \frac{1}{2\sqrt{2}} \frac{f_gL}{\sqrt{N}} \left|{\rm sinc}\frac{f_gL\omega}{2N} \right|.
\end{equation}

Eqs. (\ref{eq_spwin_eq}) and (\ref{eq_spwinfg}) then prove Eq. (\ref{eq_spwin_main}).

We now describe the behavior of the spectral window (Eq. \ref{eq_spwin_main}) far from its main peak, in order to investigate how the total masked fraction $f_g$ and the number of holes $N$ affect its behavior. This regime is of interest in the case of MICROSCOPE.

Let us assume $\omega = \lambda \omega_s$, where $\omega_s=1/L$ is the sampling frequency, with $\lambda=O(1)$.
Then
\begin{equation}
\tilde{M}(\omega)=\frac{\sqrt{N}}{\lambda\omega_s\sqrt{2}} \sin\frac{f_g\lambda}{2N}.
\end{equation}

Therefore, the spectral window is bounded by $b = \frac{\sqrt{N}}{\lambda\omega_s\sqrt{2}}$,
which does not depend on the total masked fraction, and depends only the total number of holes. At frequencies around the sampling frequency, the spectral window therefore scales as $\sqrt{N}$. 

The bound that we have just obtained, proportional to $\sqrt{N}$ is true for every frequency (or equivalently, for every $\lambda$), becoming $b(\omega) =\frac{\sqrt{N}}{\omega\sqrt{2}}$.

\subsection{Numerical example} \label{sect_numexp}

To illustrate the behavior of the spectral window, we apply three classes of gaps to the expected MICROSCOPE differential acceleration noise. Those classes allow us to compare the effect of both the total fraction of masked data and the number of holes, and are defined as:

\begin{itemize}
\item Class A: 26,000 holes represent 20\% of the total signal, with a size of 20 data points
\item Class B: 2,600 holes represent 20\% of the total signal, with a size of 200 data points
\item Class C: 2,600 holes represent 2\% of the total signal, with a size of 20 data points
\end{itemize}

\begin{figure}
 \includegraphics[width=0.45\textwidth]{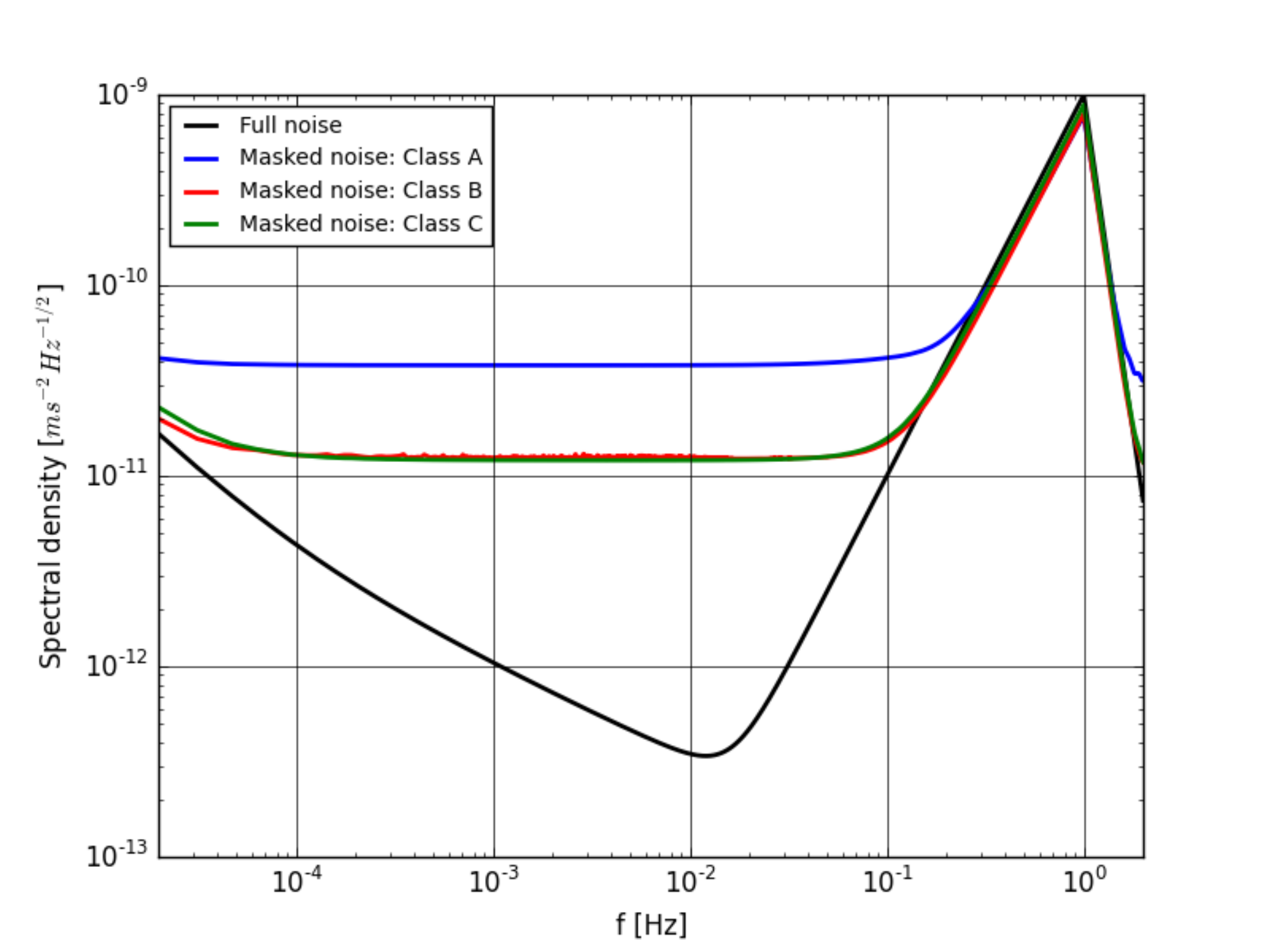}
\caption{Spectral leakage affecting the MICROSCOPE nominal differential acceleration's noise PSD (shown here is its square root) due to missing data. The black line shows the noise with no missing value, while the colored curves show the effect of missing data, depending on the geometry of the mask (classes A, B and C are shown in blue, green and red, respectively). The green curve completely overlaps the red one for frequencies less than 0.1 Hz.}
\label{fig_leakage}       % Give a unique label
\end{figure}

The observed effect for each class of gaps is shown as the square root of the PSD in Fig. \ref{fig_leakage}.
The spectral leakage originates from the frequencies where the PSD is highest: mostly 1 Hz, but some spectral leakage from the lowest frequencies is also visible. 
Far from the frequencies from where the leakage occurs, we can observe that Classes B (red) and C (green) produce a similar spectral leakage, $\sqrt{10}$ lower than that of Class A (which has 10 times as many gaps than Classes B and C). It illustrates that far from the origin of the leakage, the spectral leakage due to missing data depends only on the number of holes, as shown analytically above.

\subsection{Discussion} \label{sect_discussion}

We conclude this Appendix with a short summary of the behavior of the spectral window corresponding to randomly distributed gaps (Eq. \ref{eq_spwin_main}) that go beyond our limited MICROSCOPE application:
 
\begin{itemize}
\item the spectral window at $\omega = 0$ depends on the total masked fraction $f_g$ only; this corresponds to the loss of power due to gaps
\item the behavior of the spectral leakage (linked to the spectral window at $\omega \neq 0$) at $\omega \ll 1$ (and $\omega \neq 0$) is governed by the interplay of the total masked fraction $f_g$ and of the total number of gaps $N$
\item the spectral leakage's envelope's level depends only on the total number of holes $N$ for frequencies $\omega = O(1)$ and $\omega \gg 1$.
\end{itemize}

\bibliography{micinpainting_rvt}% Produces the bibliography via BibTeX.

\providecommand{\noopsort}[1]{}\providecommand{\singleletter}[1]{#1}%
\begin{thebibliography}{35}%
\makeatletter
\providecommand \@ifxundefined [1]{%
 \@ifx{#1\undefined}
}%
\providecommand \@ifnum [1]{%
 \ifnum #1\expandafter \@firstoftwo
 \else \expandafter \@secondoftwo
 \fi
}%
\providecommand \@ifx [1]{%
 \ifx #1\expandafter \@firstoftwo
 \else \expandafter \@secondoftwo
 \fi
}%
\providecommand \natexlab [1]{#1}%
\providecommand \enquote  [1]{``#1''}%
\providecommand \bibnamefont  [1]{#1}%
\providecommand \bibfnamefont [1]{#1}%
\providecommand \citenamefont [1]{#1}%
\providecommand \href@noop [0]{\@secondoftwo}%
\providecommand \href [0]{\begingroup \@sanitize@url \@href}%
\providecommand \@href[1]{\@@startlink{#1}\@@href}%
\providecommand \@@href[1]{\endgroup#1\@@endlink}%
\providecommand \@sanitize@url [0]{\catcode `\\12\catcode `\$12\catcode
  `\&12\catcode `\#12\catcode `\^12\catcode `\_12\catcode `\%12\relax}%
\providecommand \@@startlink[1]{}%
\providecommand \@@endlink[0]{}%
\providecommand \url  [0]{\begingroup\@sanitize@url \@url }%
\providecommand \@url [1]{\endgroup\@href {#1}{\urlprefix }}%
\providecommand \urlprefix  [0]{URL }%
\providecommand \Eprint [0]{\href }%
\providecommand \doibase [0]{http://dx.doi.org/}%
\providecommand \selectlanguage [0]{\@gobble}%
\providecommand \bibinfo  [0]{\@secondoftwo}%
\providecommand \bibfield  [0]{\@secondoftwo}%
\providecommand \translation [1]{[#1]}%
\providecommand \BibitemOpen [0]{}%
\providecommand \bibitemStop [0]{}%
\providecommand \bibitemNoStop [0]{.\EOS\space}%
\providecommand \EOS [0]{\spacefactor3000\relax}%
\providecommand \BibitemShut  [1]{\csname bibitem#1\endcsname}%
\let\auto@bib@innerbib\@empty
%</preamble>
\bibitem [{\citenamefont {{de Jong}}\ \emph {et~al.}(2015)\citenamefont {{de
  Jong}}, \citenamefont {{Verdoes Kleijn}}, \citenamefont {{Boxhoorn}},
  \citenamefont {{Buddelmeijer}}, \citenamefont {{Capaccioli}}, \citenamefont
  {{Getman}}, \citenamefont {{Grado}}, \citenamefont {{Helmich}}, \citenamefont
  {{Huang}}, \citenamefont {{Irisarri}}, \citenamefont {{Kuijken}},
  \citenamefont {{LaBarbera}}, \citenamefont {{McFarland}}, \citenamefont
  {{Napolitano}}, \citenamefont {{Radovich}}, \citenamefont {{Sikkema}},
  \citenamefont {{Valentijn}}, \citenamefont {{Begeman}}, \citenamefont
  {{Brescia}}, \citenamefont {{Cavuoti}}, \citenamefont {{Choi}}, \citenamefont
  {{Cordes}}, \citenamefont {{Covone}}, \citenamefont {{Dall'Ora}},
  \citenamefont {{Hildebrandt}}, \citenamefont {{Longo}}, \citenamefont
  {{Nakajima}}, \citenamefont {{Paolillo}}, \citenamefont {{Puddu}},
  \citenamefont {{Rifatto}}, \citenamefont {{Tortora}}, \citenamefont {{van
  Uitert}}, \citenamefont {{Buddendiek}}, \citenamefont {{Harnois-D{\'e}raps}},
  \citenamefont {{Erben}}, \citenamefont {{Eriksen}}, \citenamefont
  {{Heymans}}, \citenamefont {{Hoekstra}}, \citenamefont {{Joachimi}},
  \citenamefont {{Kitching}}, \citenamefont {{Klaes}}, \citenamefont
  {{Koopmans}}, \citenamefont {{K{\"o}hlinger}}, \citenamefont {{Roy}},
  \citenamefont {{Sifon}}, \citenamefont {{Schneider}}, \citenamefont
  {{Sutherland}}, \citenamefont {{Viola}},\ and\ \citenamefont
  {{Vriend}}}]{dejong15}%
  \BibitemOpen
  \bibfield  {author} {\bibinfo {author} {\bibfnamefont {J.~T.~A.}\
  \bibnamefont {{de Jong}}}, \bibinfo {author} {\bibfnamefont {G.~A.}\
  \bibnamefont {{Verdoes Kleijn}}}, \bibinfo {author} {\bibfnamefont {D.~R.}\
  \bibnamefont {{Boxhoorn}}}, \bibinfo {author} {\bibfnamefont
  {H.}~\bibnamefont {{Buddelmeijer}}}, \bibinfo {author} {\bibfnamefont
  {M.}~\bibnamefont {{Capaccioli}}}, \bibinfo {author} {\bibfnamefont
  {F.}~\bibnamefont {{Getman}}}, \bibinfo {author} {\bibfnamefont
  {A.}~\bibnamefont {{Grado}}}, \bibinfo {author} {\bibfnamefont
  {E.}~\bibnamefont {{Helmich}}}, \bibinfo {author} {\bibfnamefont
  {Z.}~\bibnamefont {{Huang}}}, \bibinfo {author} {\bibfnamefont
  {N.}~\bibnamefont {{Irisarri}}}, \bibinfo {author} {\bibfnamefont
  {K.}~\bibnamefont {{Kuijken}}}, \bibinfo {author} {\bibfnamefont
  {F.}~\bibnamefont {{LaBarbera}}}, \bibinfo {author} {\bibfnamefont {J.~P.}\
  \bibnamefont {{McFarland}}}, \bibinfo {author} {\bibfnamefont {N.~R.}\
  \bibnamefont {{Napolitano}}}, \bibinfo {author} {\bibfnamefont
  {M.}~\bibnamefont {{Radovich}}}, \bibinfo {author} {\bibfnamefont
  {G.}~\bibnamefont {{Sikkema}}}, \bibinfo {author} {\bibfnamefont {E.~A.}\
  \bibnamefont {{Valentijn}}}, \bibinfo {author} {\bibfnamefont {K.~G.}\
  \bibnamefont {{Begeman}}}, \bibinfo {author} {\bibfnamefont {M.}~\bibnamefont
  {{Brescia}}}, \bibinfo {author} {\bibfnamefont {S.}~\bibnamefont
  {{Cavuoti}}}, \bibinfo {author} {\bibfnamefont {A.}~\bibnamefont {{Choi}}},
  \bibinfo {author} {\bibfnamefont {O.-M.}\ \bibnamefont {{Cordes}}}, \bibinfo
  {author} {\bibfnamefont {G.}~\bibnamefont {{Covone}}}, \bibinfo {author}
  {\bibfnamefont {M.}~\bibnamefont {{Dall'Ora}}}, \bibinfo {author}
  {\bibfnamefont {H.}~\bibnamefont {{Hildebrandt}}}, \bibinfo {author}
  {\bibfnamefont {G.}~\bibnamefont {{Longo}}}, \bibinfo {author} {\bibfnamefont
  {R.}~\bibnamefont {{Nakajima}}}, \bibinfo {author} {\bibfnamefont
  {M.}~\bibnamefont {{Paolillo}}}, \bibinfo {author} {\bibfnamefont
  {E.}~\bibnamefont {{Puddu}}}, \bibinfo {author} {\bibfnamefont
  {A.}~\bibnamefont {{Rifatto}}}, \bibinfo {author} {\bibfnamefont
  {C.}~\bibnamefont {{Tortora}}}, \bibinfo {author} {\bibfnamefont
  {E.}~\bibnamefont {{van Uitert}}}, \bibinfo {author} {\bibfnamefont
  {A.}~\bibnamefont {{Buddendiek}}}, \bibinfo {author} {\bibfnamefont
  {J.}~\bibnamefont {{Harnois-D{\'e}raps}}}, \bibinfo {author} {\bibfnamefont
  {T.}~\bibnamefont {{Erben}}}, \bibinfo {author} {\bibfnamefont {M.~B.}\
  \bibnamefont {{Eriksen}}}, \bibinfo {author} {\bibfnamefont {C.}~\bibnamefont
  {{Heymans}}}, \bibinfo {author} {\bibfnamefont {H.}~\bibnamefont
  {{Hoekstra}}}, \bibinfo {author} {\bibfnamefont {B.}~\bibnamefont
  {{Joachimi}}}, \bibinfo {author} {\bibfnamefont {T.~D.}\ \bibnamefont
  {{Kitching}}}, \bibinfo {author} {\bibfnamefont {D.}~\bibnamefont {{Klaes}}},
  \bibinfo {author} {\bibfnamefont {L.~V.~E.}\ \bibnamefont {{Koopmans}}},
  \bibinfo {author} {\bibfnamefont {F.}~\bibnamefont {{K{\"o}hlinger}}},
  \bibinfo {author} {\bibfnamefont {N.}~\bibnamefont {{Roy}}}, \bibinfo
  {author} {\bibfnamefont {C.}~\bibnamefont {{Sifon}}}, \bibinfo {author}
  {\bibfnamefont {P.}~\bibnamefont {{Schneider}}}, \bibinfo {author}
  {\bibfnamefont {W.~J.}\ \bibnamefont {{Sutherland}}}, \bibinfo {author}
  {\bibfnamefont {M.}~\bibnamefont {{Viola}}}, \ and\ \bibinfo {author}
  {\bibfnamefont {W.-J.}\ \bibnamefont {{Vriend}}},\ }\href@noop {} {\bibfield
  {journal} {\bibinfo  {journal} {ArXiv e-prints}\ } (\bibinfo {year}
  {2015})},\ \Eprint {http://arxiv.org/abs/1507.00742} {arXiv:1507.00742}
  \BibitemShut {NoStop}%
\bibitem [{\citenamefont {{Vallisneri}}\ \emph {et~al.}(2015)\citenamefont
  {{Vallisneri}}, \citenamefont {{Kanner}}, \citenamefont {{Williams}},
  \citenamefont {{Weinstein}},\ and\ \citenamefont
  {{Stephens}}}]{vallisneri15}%
  \BibitemOpen
  \bibfield  {author} {\bibinfo {author} {\bibfnamefont {M.}~\bibnamefont
  {{Vallisneri}}}, \bibinfo {author} {\bibfnamefont {J.}~\bibnamefont
  {{Kanner}}}, \bibinfo {author} {\bibfnamefont {R.}~\bibnamefont
  {{Williams}}}, \bibinfo {author} {\bibfnamefont {A.}~\bibnamefont
  {{Weinstein}}}, \ and\ \bibinfo {author} {\bibfnamefont {B.}~\bibnamefont
  {{Stephens}}},\ }\href {\doibase 10.1088/1742-6596/610/1/012021} {\bibfield
  {journal} {\bibinfo  {journal} {Journal of Physics Conference Series}\
  }\textbf {\bibinfo {volume} {610}},\ \bibinfo {eid} {012021} (\bibinfo {year}
  {2015})},\ \Eprint {http://arxiv.org/abs/1410.4839} {arXiv:1410.4839 [gr-qc]}
  \BibitemShut {NoStop}%
\bibitem [{\citenamefont {{Starck}}\ \emph {et~al.}(2013)\citenamefont
  {{Starck}}, \citenamefont {{Donoho}}, \citenamefont {{Fadili}},\ and\
  \citenamefont {{Rassat}}}]{starck13}%
  \BibitemOpen
  \bibfield  {author} {\bibinfo {author} {\bibfnamefont {J.-L.}\ \bibnamefont
  {{Starck}}}, \bibinfo {author} {\bibfnamefont {D.~L.}\ \bibnamefont
  {{Donoho}}}, \bibinfo {author} {\bibfnamefont {M.~J.}\ \bibnamefont
  {{Fadili}}}, \ and\ \bibinfo {author} {\bibfnamefont {A.}~\bibnamefont
  {{Rassat}}},\ }\href {\doibase 10.1051/0004-6361/201321257} {\bibfield
  {journal} {\bibinfo  {journal} {\aap}\ }\textbf {\bibinfo {volume} {552}},\
  \bibinfo {eid} {A133} (\bibinfo {year} {2013})},\ \Eprint
  {http://arxiv.org/abs/1302.2758} {arXiv:1302.2758 [astro-ph.IM]} \BibitemShut
  {NoStop}%
\bibitem [{\citenamefont {{Planck Collaboration}}\ \emph
  {et~al.}(2014)\citenamefont {{Planck Collaboration}}, \citenamefont {{Ade}},
  \citenamefont {{Aghanim}}, \citenamefont {{Alves}}, \citenamefont
  {{Armitage-Caplan}}, \citenamefont {{Arnaud}}, \citenamefont {{Ashdown}},
  \citenamefont {{Atrio-Barandela}}, \citenamefont {{Aumont}}, \citenamefont
  {{Aussel}},\ and\ \citenamefont {et~al.}}]{planck14}%
  \BibitemOpen
  \bibfield  {author} {\bibinfo {author} {\bibnamefont {{Planck
  Collaboration}}}, \bibinfo {author} {\bibfnamefont {P.~A.~R.}\ \bibnamefont
  {{Ade}}}, \bibinfo {author} {\bibfnamefont {N.}~\bibnamefont {{Aghanim}}},
  \bibinfo {author} {\bibfnamefont {M.~I.~R.}\ \bibnamefont {{Alves}}},
  \bibinfo {author} {\bibfnamefont {C.}~\bibnamefont {{Armitage-Caplan}}},
  \bibinfo {author} {\bibfnamefont {M.}~\bibnamefont {{Arnaud}}}, \bibinfo
  {author} {\bibfnamefont {M.}~\bibnamefont {{Ashdown}}}, \bibinfo {author}
  {\bibfnamefont {F.}~\bibnamefont {{Atrio-Barandela}}}, \bibinfo {author}
  {\bibfnamefont {J.}~\bibnamefont {{Aumont}}}, \bibinfo {author}
  {\bibfnamefont {H.}~\bibnamefont {{Aussel}}}, \ and\ \bibinfo {author}
  {\bibnamefont {et~al.}},\ }\href {\doibase 10.1051/0004-6361/201321529}
  {\bibfield  {journal} {\bibinfo  {journal} {\aap}\ }\textbf {\bibinfo
  {volume} {571}},\ \bibinfo {eid} {A1} (\bibinfo {year} {2014})},\ \Eprint
  {http://arxiv.org/abs/1303.5062} {arXiv:1303.5062} \BibitemShut {NoStop}%
\bibitem [{\citenamefont {{Rassat}}\ \emph {et~al.}(2014)\citenamefont
  {{Rassat}}, \citenamefont {{Starck}}, \citenamefont {{Paykari}},
  \citenamefont {{Sureau}},\ and\ \citenamefont {{Bobin}}}]{rassat14}%
  \BibitemOpen
  \bibfield  {author} {\bibinfo {author} {\bibfnamefont {A.}~\bibnamefont
  {{Rassat}}}, \bibinfo {author} {\bibfnamefont {J.-L.}\ \bibnamefont
  {{Starck}}}, \bibinfo {author} {\bibfnamefont {P.}~\bibnamefont {{Paykari}}},
  \bibinfo {author} {\bibfnamefont {F.}~\bibnamefont {{Sureau}}}, \ and\
  \bibinfo {author} {\bibfnamefont {J.}~\bibnamefont {{Bobin}}},\ }\href
  {\doibase 10.1088/1475-7516/2014/08/006} {\bibfield  {journal} {\bibinfo
  {journal} {\jcap}\ }\textbf {\bibinfo {volume} {8}},\ \bibinfo {eid} {006}
  (\bibinfo {year} {2014})},\ \Eprint {http://arxiv.org/abs/1405.1844}
  {arXiv:1405.1844} \BibitemShut {NoStop}%
\bibitem [{\citenamefont {{Pires}}\ \emph {et~al.}(2009)\citenamefont
  {{Pires}}, \citenamefont {{Starck}}, \citenamefont {{Amara}}, \citenamefont
  {{Teyssier}}, \citenamefont {{R{\'e}fr{\'e}gier}},\ and\ \citenamefont
  {{Fadili}}}]{pires09}%
  \BibitemOpen
  \bibfield  {author} {\bibinfo {author} {\bibfnamefont {S.}~\bibnamefont
  {{Pires}}}, \bibinfo {author} {\bibfnamefont {J.-L.}\ \bibnamefont
  {{Starck}}}, \bibinfo {author} {\bibfnamefont {A.}~\bibnamefont {{Amara}}},
  \bibinfo {author} {\bibfnamefont {R.}~\bibnamefont {{Teyssier}}}, \bibinfo
  {author} {\bibfnamefont {A.}~\bibnamefont {{R{\'e}fr{\'e}gier}}}, \ and\
  \bibinfo {author} {\bibfnamefont {J.}~\bibnamefont {{Fadili}}},\ }\href
  {\doibase 10.1111/j.1365-2966.2009.14625.x} {\bibfield  {journal} {\bibinfo
  {journal} {\mnras}\ }\textbf {\bibinfo {volume} {395}},\ \bibinfo {pages}
  {1265} (\bibinfo {year} {2009})},\ \Eprint {http://arxiv.org/abs/0804.4068}
  {arXiv:0804.4068} \BibitemShut {NoStop}%
\bibitem [{\citenamefont {{Liu}}\ \emph {et~al.}(2014)\citenamefont {{Liu}},
  \citenamefont {{Wang}}, \citenamefont {{Pan}},\ and\ \citenamefont
  {{Fan}}}]{liu14}%
  \BibitemOpen
  \bibfield  {author} {\bibinfo {author} {\bibfnamefont {X.}~\bibnamefont
  {{Liu}}}, \bibinfo {author} {\bibfnamefont {Q.}~\bibnamefont {{Wang}}},
  \bibinfo {author} {\bibfnamefont {C.}~\bibnamefont {{Pan}}}, \ and\ \bibinfo
  {author} {\bibfnamefont {Z.}~\bibnamefont {{Fan}}},\ }\href {\doibase
  10.1088/0004-637X/784/1/31} {\bibfield  {journal} {\bibinfo  {journal}
  {\apj}\ }\textbf {\bibinfo {volume} {784}},\ \bibinfo {eid} {31} (\bibinfo
  {year} {2014})},\ \Eprint {http://arxiv.org/abs/1304.2873} {arXiv:1304.2873}
  \BibitemShut {NoStop}%
\bibitem [{\citenamefont {{Shirasaki}}\ \emph {et~al.}(2013)\citenamefont
  {{Shirasaki}}, \citenamefont {{Yoshida}},\ and\ \citenamefont
  {{Hamana}}}]{shirasaki14}%
  \BibitemOpen
  \bibfield  {author} {\bibinfo {author} {\bibfnamefont {M.}~\bibnamefont
  {{Shirasaki}}}, \bibinfo {author} {\bibfnamefont {N.}~\bibnamefont
  {{Yoshida}}}, \ and\ \bibinfo {author} {\bibfnamefont {T.}~\bibnamefont
  {{Hamana}}},\ }\href {\doibase 10.1088/0004-637X/774/2/111} {\bibfield
  {journal} {\bibinfo  {journal} {\apj}\ }\textbf {\bibinfo {volume} {774}},\
  \bibinfo {eid} {111} (\bibinfo {year} {2013})},\ \Eprint
  {http://arxiv.org/abs/1304.2164} {arXiv:1304.2164} \BibitemShut {NoStop}%
\bibitem [{\citenamefont {{Garc{\'{\i}}a}}\ \emph {et~al.}(2014)\citenamefont
  {{Garc{\'{\i}}a}}, \citenamefont {{Mathur}}, \citenamefont {{Pires}},
  \citenamefont {{R{\'e}gulo}}, \citenamefont {{Bellamy}}, \citenamefont
  {{Pall{\'e}}}, \citenamefont {{Ballot}}, \citenamefont {{Barcel{\'o}
  Forteza}}, \citenamefont {{Beck}}, \citenamefont {{Bedding}}, \citenamefont
  {{Ceillier}}, \citenamefont {{Roca Cort{\'e}s}}, \citenamefont {{Salabert}},\
  and\ \citenamefont {{Stello}}}]{garcia14}%
  \BibitemOpen
  \bibfield  {author} {\bibinfo {author} {\bibfnamefont {R.~A.}\ \bibnamefont
  {{Garc{\'{\i}}a}}}, \bibinfo {author} {\bibfnamefont {S.}~\bibnamefont
  {{Mathur}}}, \bibinfo {author} {\bibfnamefont {S.}~\bibnamefont {{Pires}}},
  \bibinfo {author} {\bibfnamefont {C.}~\bibnamefont {{R{\'e}gulo}}}, \bibinfo
  {author} {\bibfnamefont {B.}~\bibnamefont {{Bellamy}}}, \bibinfo {author}
  {\bibfnamefont {P.~L.}\ \bibnamefont {{Pall{\'e}}}}, \bibinfo {author}
  {\bibfnamefont {J.}~\bibnamefont {{Ballot}}}, \bibinfo {author}
  {\bibfnamefont {S.}~\bibnamefont {{Barcel{\'o} Forteza}}}, \bibinfo {author}
  {\bibfnamefont {P.~G.}\ \bibnamefont {{Beck}}}, \bibinfo {author}
  {\bibfnamefont {T.~R.}\ \bibnamefont {{Bedding}}}, \bibinfo {author}
  {\bibfnamefont {T.}~\bibnamefont {{Ceillier}}}, \bibinfo {author}
  {\bibfnamefont {T.}~\bibnamefont {{Roca Cort{\'e}s}}}, \bibinfo {author}
  {\bibfnamefont {D.}~\bibnamefont {{Salabert}}}, \ and\ \bibinfo {author}
  {\bibfnamefont {D.}~\bibnamefont {{Stello}}},\ }\href {\doibase
  10.1051/0004-6361/201323326} {\bibfield  {journal} {\bibinfo  {journal}
  {\aap}\ }\textbf {\bibinfo {volume} {568}},\ \bibinfo {eid} {A10} (\bibinfo
  {year} {2014})},\ \Eprint {http://arxiv.org/abs/1405.5374} {arXiv:1405.5374
  [astro-ph.SR]} \BibitemShut {NoStop}%
\bibitem [{\citenamefont {{Pires}}\ \emph {et~al.}(2015)\citenamefont
  {{Pires}}, \citenamefont {{Mathur}}, \citenamefont {{Garc{\'{\i}}a}},
  \citenamefont {{Ballot}}, \citenamefont {{Stello}},\ and\ \citenamefont
  {{Sato}}}]{pires15}%
  \BibitemOpen
  \bibfield  {author} {\bibinfo {author} {\bibfnamefont {S.}~\bibnamefont
  {{Pires}}}, \bibinfo {author} {\bibfnamefont {S.}~\bibnamefont {{Mathur}}},
  \bibinfo {author} {\bibfnamefont {R.~A.}\ \bibnamefont {{Garc{\'{\i}}a}}},
  \bibinfo {author} {\bibfnamefont {J.}~\bibnamefont {{Ballot}}}, \bibinfo
  {author} {\bibfnamefont {D.}~\bibnamefont {{Stello}}}, \ and\ \bibinfo
  {author} {\bibfnamefont {K.}~\bibnamefont {{Sato}}},\ }\href {\doibase
  10.1051/0004-6361/201322361} {\bibfield  {journal} {\bibinfo  {journal}
  {Astronomy and Astrophysics}\ }\textbf {\bibinfo {volume} {574}},\ \bibinfo
  {eid} {A18} (\bibinfo {year} {2015})},\ \Eprint
  {http://arxiv.org/abs/1410.6088} {arXiv:1410.6088 [astro-ph.SR]} \BibitemShut
  {NoStop}%
\bibitem [{\citenamefont {{Baghi}}\ \emph {et~al.}(2015)\citenamefont
  {{Baghi}}, \citenamefont {{M{\'e}tris}}, \citenamefont {{Berg{\'e}}},
  \citenamefont {{Christophe}}, \citenamefont {{Touboul}},\ and\ \citenamefont
  {{Rodrigues}}}]{baghi15}%
  \BibitemOpen
  \bibfield  {author} {\bibinfo {author} {\bibfnamefont {Q.}~\bibnamefont
  {{Baghi}}}, \bibinfo {author} {\bibfnamefont {G.}~\bibnamefont
  {{M{\'e}tris}}}, \bibinfo {author} {\bibfnamefont {J.}~\bibnamefont
  {{Berg{\'e}}}}, \bibinfo {author} {\bibfnamefont {B.}~\bibnamefont
  {{Christophe}}}, \bibinfo {author} {\bibfnamefont {P.}~\bibnamefont
  {{Touboul}}}, \ and\ \bibinfo {author} {\bibfnamefont {M.}~\bibnamefont
  {{Rodrigues}}},\ }\href {\doibase 10.1103/PhysRevD.91.062003} {\bibfield
  {journal} {\bibinfo  {journal} {\prd}\ }\textbf {\bibinfo {volume} {91}},\
  \bibinfo {eid} {062003} (\bibinfo {year} {2015})},\ \Eprint
  {http://arxiv.org/abs/1503.01470} {arXiv:1503.01470 [gr-qc]} \BibitemShut
  {NoStop}%
\bibitem [{\citenamefont {{Touboul}}\ and\ \citenamefont
  {{Rodrigues}}(2001)}]{touboul01}%
  \BibitemOpen
  \bibfield  {author} {\bibinfo {author} {\bibfnamefont {P.}~\bibnamefont
  {{Touboul}}}\ and\ \bibinfo {author} {\bibfnamefont {M.}~\bibnamefont
  {{Rodrigues}}},\ }\href {\doibase 10.1088/0264-9381/18/13/311} {\bibfield
  {journal} {\bibinfo  {journal} {Classical and Quantum Gravity}\ }\textbf
  {\bibinfo {volume} {18}},\ \bibinfo {pages} {2487} (\bibinfo {year}
  {2001})}\BibitemShut {NoStop}%
\bibitem [{\citenamefont {{Touboul}}(2009)}]{touboul09}%
  \BibitemOpen
  \bibfield  {author} {\bibinfo {author} {\bibfnamefont {P.}~\bibnamefont
  {{Touboul}}},\ }\href {\doibase 10.1007/s11214-009-9565-y} {\bibfield
  {journal} {\bibinfo  {journal} {\ssr}\ }\textbf {\bibinfo {volume} {148}},\
  \bibinfo {pages} {455} (\bibinfo {year} {2009})}\BibitemShut {NoStop}%
\bibitem [{\citenamefont {{Touboul}}\ \emph {et~al.}(2012)\citenamefont
  {{Touboul}}, \citenamefont {{M{\'e}tris}}, \citenamefont {{Lebat}},\ and\
  \citenamefont {{Robert}}}]{touboul12}%
  \BibitemOpen
  \bibfield  {author} {\bibinfo {author} {\bibfnamefont {P.}~\bibnamefont
  {{Touboul}}}, \bibinfo {author} {\bibfnamefont {G.}~\bibnamefont
  {{M{\'e}tris}}}, \bibinfo {author} {\bibfnamefont {V.}~\bibnamefont
  {{Lebat}}}, \ and\ \bibinfo {author} {\bibfnamefont {A.}~\bibnamefont
  {{Robert}}},\ }\href {\doibase 10.1088/0264-9381/29/18/184010} {\bibfield
  {journal} {\bibinfo  {journal} {Classical and Quantum Gravity}\ }\textbf
  {\bibinfo {volume} {29}},\ \bibinfo {eid} {184010} (\bibinfo {year}
  {2012})}\BibitemShut {NoStop}%
\bibitem [{\citenamefont {{Berg{\'e}}}\ \emph {et~al.}(2015)\citenamefont
  {{Berg{\'e}}}, \citenamefont {{Touboul}},\ and\ \citenamefont
  {{Rodrigues}}}]{berge15a}%
  \BibitemOpen
  \bibfield  {author} {\bibinfo {author} {\bibfnamefont {J.}~\bibnamefont
  {{Berg{\'e}}}}, \bibinfo {author} {\bibfnamefont {P.}~\bibnamefont
  {{Touboul}}}, \ and\ \bibinfo {author} {\bibfnamefont {M.}~\bibnamefont
  {{Rodrigues}}},\ }\href {\doibase 10.1088/1742-6596/610/1/012009} {\bibfield
  {journal} {\bibinfo  {journal} {Journal of Physics Conference Series}\
  }\textbf {\bibinfo {volume} {610}},\ \bibinfo {eid} {012009} (\bibinfo {year}
  {2015})},\ \Eprint {http://arxiv.org/abs/1501.01644} {arXiv:1501.01644
  [gr-qc]} \BibitemShut {NoStop}%
\bibitem [{\citenamefont {{McNamara}}\ \emph {et~al.}(2008)\citenamefont
  {{McNamara}}, \citenamefont {{Vitale}}, \citenamefont {{Danzmann}},\ and\
  \citenamefont {{LISA Pathfinder Science Working Team}}}]{mcnamara08}%
  \BibitemOpen
  \bibfield  {author} {\bibinfo {author} {\bibfnamefont {P.}~\bibnamefont
  {{McNamara}}}, \bibinfo {author} {\bibfnamefont {S.}~\bibnamefont
  {{Vitale}}}, \bibinfo {author} {\bibfnamefont {K.}~\bibnamefont
  {{Danzmann}}}, \ and\ \bibinfo {author} {\bibnamefont {{LISA Pathfinder
  Science Working Team}}},\ }\href {\doibase 10.1088/0264-9381/25/11/114034}
  {\bibfield  {journal} {\bibinfo  {journal} {Classical and Quantum Gravity}\
  }\textbf {\bibinfo {volume} {25}},\ \bibinfo {eid} {114034} (\bibinfo {year}
  {2008})}\BibitemShut {NoStop}%
\bibitem [{\citenamefont {{Armano}}\ \emph {et~al.}(2009)\citenamefont
  {{Armano}}, \citenamefont {{Benedetti}}, \citenamefont {{Bogenstahl}},
  \citenamefont {{Bortoluzzi}}, \citenamefont {{Bosetti}}, \citenamefont
  {{Brandt}}, \citenamefont {{Cavalleri}}, \citenamefont {{Ciani}},
  \citenamefont {{Cristofolini}}, \citenamefont {{Cruise}}, \citenamefont
  {{Danzmann}}, \citenamefont {{Diepholz}}, \citenamefont {{Dixon}},
  \citenamefont {{Dolesi}}, \citenamefont {{Fauste}}, \citenamefont
  {{Ferraioli}}, \citenamefont {{Fertin}}, \citenamefont {{Fichter}},
  \citenamefont {{Freschi}}, \citenamefont {{Garc{\'{\i}}a}}, \citenamefont
  {{Garc{\'{\i}}a}}, \citenamefont {{Grynagier}}, \citenamefont {{Guzm{\'a}n}},
  \citenamefont {{Fitzsimons}}, \citenamefont {{Heinzel}}, \citenamefont
  {{Hewitson}}, \citenamefont {{Hollington}}, \citenamefont {{Hough}},
  \citenamefont {{Hueller}}, \citenamefont {{Hoyland}}, \citenamefont
  {{Jennrich}}, \citenamefont {{Johlander}}, \citenamefont {{Killow}},
  \citenamefont {{Lobo}}, \citenamefont {{Mance}}, \citenamefont {{Mateos}},
  \citenamefont {{McNamara}}, \citenamefont {{Monsky}}, \citenamefont
  {{Nicolini}}, \citenamefont {{Nicolodi}}, \citenamefont {{Nofrarias}},
  \citenamefont {{Perreur-Lloyd}}, \citenamefont {{Plagnol}}, \citenamefont
  {{Racca}}, \citenamefont {{Ramos-Castro}}, \citenamefont {{Robertson}},
  \citenamefont {{Sanjuan}}, \citenamefont {{Schulte}}, \citenamefont
  {{Shaul}}, \citenamefont {{Smit}}, \citenamefont {{Stagnaro}}, \citenamefont
  {{Steier}}, \citenamefont {{Sumner}}, \citenamefont {{Tateo}}, \citenamefont
  {{Tombolato}}, \citenamefont {{Vischer}}, \citenamefont {{Vitale}},
  \citenamefont {{Wanner}}, \citenamefont {{Ward}}, \citenamefont {{Waschke}},
  \citenamefont {{Wand}}, \citenamefont {{Wass}}, \citenamefont {{Weber}},
  \citenamefont {{Ziegler}},\ and\ \citenamefont {{Zweifel}}}]{armano09}%
  \BibitemOpen
  \bibfield  {author} {\bibinfo {author} {\bibfnamefont {M.}~\bibnamefont
  {{Armano}}}, \bibinfo {author} {\bibfnamefont {M.}~\bibnamefont
  {{Benedetti}}}, \bibinfo {author} {\bibfnamefont {J.}~\bibnamefont
  {{Bogenstahl}}}, \bibinfo {author} {\bibfnamefont {D.}~\bibnamefont
  {{Bortoluzzi}}}, \bibinfo {author} {\bibfnamefont {P.}~\bibnamefont
  {{Bosetti}}}, \bibinfo {author} {\bibfnamefont {N.}~\bibnamefont {{Brandt}}},
  \bibinfo {author} {\bibfnamefont {A.}~\bibnamefont {{Cavalleri}}}, \bibinfo
  {author} {\bibfnamefont {G.}~\bibnamefont {{Ciani}}}, \bibinfo {author}
  {\bibfnamefont {I.}~\bibnamefont {{Cristofolini}}}, \bibinfo {author}
  {\bibfnamefont {A.~M.}\ \bibnamefont {{Cruise}}}, \bibinfo {author}
  {\bibfnamefont {K.}~\bibnamefont {{Danzmann}}}, \bibinfo {author}
  {\bibfnamefont {I.}~\bibnamefont {{Diepholz}}}, \bibinfo {author}
  {\bibfnamefont {G.}~\bibnamefont {{Dixon}}}, \bibinfo {author} {\bibfnamefont
  {R.}~\bibnamefont {{Dolesi}}}, \bibinfo {author} {\bibfnamefont
  {J.}~\bibnamefont {{Fauste}}}, \bibinfo {author} {\bibfnamefont
  {L.}~\bibnamefont {{Ferraioli}}}, \bibinfo {author} {\bibfnamefont
  {D.}~\bibnamefont {{Fertin}}}, \bibinfo {author} {\bibfnamefont
  {W.}~\bibnamefont {{Fichter}}}, \bibinfo {author} {\bibfnamefont
  {M.}~\bibnamefont {{Freschi}}}, \bibinfo {author} {\bibfnamefont
  {A.}~\bibnamefont {{Garc{\'{\i}}a}}}, \bibinfo {author} {\bibfnamefont
  {C.}~\bibnamefont {{Garc{\'{\i}}a}}}, \bibinfo {author} {\bibfnamefont
  {A.}~\bibnamefont {{Grynagier}}}, \bibinfo {author} {\bibfnamefont
  {F.}~\bibnamefont {{Guzm{\'a}n}}}, \bibinfo {author} {\bibfnamefont
  {E.}~\bibnamefont {{Fitzsimons}}}, \bibinfo {author} {\bibfnamefont
  {G.}~\bibnamefont {{Heinzel}}}, \bibinfo {author} {\bibfnamefont
  {M.}~\bibnamefont {{Hewitson}}}, \bibinfo {author} {\bibfnamefont
  {D.}~\bibnamefont {{Hollington}}}, \bibinfo {author} {\bibfnamefont
  {J.}~\bibnamefont {{Hough}}}, \bibinfo {author} {\bibfnamefont
  {M.}~\bibnamefont {{Hueller}}}, \bibinfo {author} {\bibfnamefont
  {D.}~\bibnamefont {{Hoyland}}}, \bibinfo {author} {\bibfnamefont
  {O.}~\bibnamefont {{Jennrich}}}, \bibinfo {author} {\bibfnamefont
  {B.}~\bibnamefont {{Johlander}}}, \bibinfo {author} {\bibfnamefont
  {C.}~\bibnamefont {{Killow}}}, \bibinfo {author} {\bibfnamefont
  {A.}~\bibnamefont {{Lobo}}}, \bibinfo {author} {\bibfnamefont
  {D.}~\bibnamefont {{Mance}}}, \bibinfo {author} {\bibfnamefont
  {I.}~\bibnamefont {{Mateos}}}, \bibinfo {author} {\bibfnamefont {P.~W.}\
  \bibnamefont {{McNamara}}}, \bibinfo {author} {\bibfnamefont
  {A.}~\bibnamefont {{Monsky}}}, \bibinfo {author} {\bibfnamefont
  {D.}~\bibnamefont {{Nicolini}}}, \bibinfo {author} {\bibfnamefont
  {D.}~\bibnamefont {{Nicolodi}}}, \bibinfo {author} {\bibfnamefont
  {M.}~\bibnamefont {{Nofrarias}}}, \bibinfo {author} {\bibfnamefont
  {M.}~\bibnamefont {{Perreur-Lloyd}}}, \bibinfo {author} {\bibfnamefont
  {E.}~\bibnamefont {{Plagnol}}}, \bibinfo {author} {\bibfnamefont {G.~D.}\
  \bibnamefont {{Racca}}}, \bibinfo {author} {\bibfnamefont {J.}~\bibnamefont
  {{Ramos-Castro}}}, \bibinfo {author} {\bibfnamefont {D.}~\bibnamefont
  {{Robertson}}}, \bibinfo {author} {\bibfnamefont {J.}~\bibnamefont
  {{Sanjuan}}}, \bibinfo {author} {\bibfnamefont {M.~O.}\ \bibnamefont
  {{Schulte}}}, \bibinfo {author} {\bibfnamefont {D.~N.~A.}\ \bibnamefont
  {{Shaul}}}, \bibinfo {author} {\bibfnamefont {M.}~\bibnamefont {{Smit}}},
  \bibinfo {author} {\bibfnamefont {L.}~\bibnamefont {{Stagnaro}}}, \bibinfo
  {author} {\bibfnamefont {F.}~\bibnamefont {{Steier}}}, \bibinfo {author}
  {\bibfnamefont {T.~J.}\ \bibnamefont {{Sumner}}}, \bibinfo {author}
  {\bibfnamefont {N.}~\bibnamefont {{Tateo}}}, \bibinfo {author} {\bibfnamefont
  {D.}~\bibnamefont {{Tombolato}}}, \bibinfo {author} {\bibfnamefont
  {G.}~\bibnamefont {{Vischer}}}, \bibinfo {author} {\bibfnamefont
  {S.}~\bibnamefont {{Vitale}}}, \bibinfo {author} {\bibfnamefont
  {G.}~\bibnamefont {{Wanner}}}, \bibinfo {author} {\bibfnamefont
  {H.}~\bibnamefont {{Ward}}}, \bibinfo {author} {\bibfnamefont
  {S.}~\bibnamefont {{Waschke}}}, \bibinfo {author} {\bibfnamefont
  {V.}~\bibnamefont {{Wand}}}, \bibinfo {author} {\bibfnamefont
  {P.}~\bibnamefont {{Wass}}}, \bibinfo {author} {\bibfnamefont {W.~J.}\
  \bibnamefont {{Weber}}}, \bibinfo {author} {\bibfnamefont {T.}~\bibnamefont
  {{Ziegler}}}, \ and\ \bibinfo {author} {\bibfnamefont {P.}~\bibnamefont
  {{Zweifel}}},\ }\href {\doibase 10.1088/0264-9381/26/9/094001} {\bibfield
  {journal} {\bibinfo  {journal} {Classical and Quantum Gravity}\ }\textbf
  {\bibinfo {volume} {26}},\ \bibinfo {eid} {094001} (\bibinfo {year}
  {2009})}\BibitemShut {NoStop}%
\bibitem [{\citenamefont {{Hikage}}\ \emph {et~al.}(2011)\citenamefont
  {{Hikage}}, \citenamefont {{Takada}}, \citenamefont {{Hamana}},\ and\
  \citenamefont {{Spergel}}}]{hikage11}%
  \BibitemOpen
  \bibfield  {author} {\bibinfo {author} {\bibfnamefont {C.}~\bibnamefont
  {{Hikage}}}, \bibinfo {author} {\bibfnamefont {M.}~\bibnamefont {{Takada}}},
  \bibinfo {author} {\bibfnamefont {T.}~\bibnamefont {{Hamana}}}, \ and\
  \bibinfo {author} {\bibfnamefont {D.}~\bibnamefont {{Spergel}}},\ }\href
  {\doibase 10.1111/j.1365-2966.2010.17886.x} {\bibfield  {journal} {\bibinfo
  {journal} {\mnras}\ }\textbf {\bibinfo {volume} {412}},\ \bibinfo {pages}
  {65} (\bibinfo {year} {2011})},\ \Eprint {http://arxiv.org/abs/1004.3542}
  {arXiv:1004.3542 [astro-ph.CO]} \BibitemShut {NoStop}%
\bibitem [{\citenamefont {{VanderPlas}}\ \emph {et~al.}(2012)\citenamefont
  {{VanderPlas}}, \citenamefont {{Connolly}}, \citenamefont {{Jain}},\ and\
  \citenamefont {{Jarvis}}}]{vanderplas12}%
  \BibitemOpen
  \bibfield  {author} {\bibinfo {author} {\bibfnamefont {J.~T.}\ \bibnamefont
  {{VanderPlas}}}, \bibinfo {author} {\bibfnamefont {A.~J.}\ \bibnamefont
  {{Connolly}}}, \bibinfo {author} {\bibfnamefont {B.}~\bibnamefont {{Jain}}},
  \ and\ \bibinfo {author} {\bibfnamefont {M.}~\bibnamefont {{Jarvis}}},\
  }\href {\doibase 10.1088/0004-637X/744/2/180} {\bibfield  {journal} {\bibinfo
   {journal} {\apj}\ }\textbf {\bibinfo {volume} {744}},\ \bibinfo {eid} {180}
  (\bibinfo {year} {2012})},\ \Eprint {http://arxiv.org/abs/1109.5175}
  {arXiv:1109.5175} \BibitemShut {NoStop}%
\bibitem [{\citenamefont {{Perotto}}\ \emph {et~al.}(2010)\citenamefont
  {{Perotto}}, \citenamefont {{Bobin}}, \citenamefont {{Plaszczynski}},
  \citenamefont {{Starck}},\ and\ \citenamefont {{Lavabre}}}]{perotto10}%
  \BibitemOpen
  \bibfield  {author} {\bibinfo {author} {\bibfnamefont {L.}~\bibnamefont
  {{Perotto}}}, \bibinfo {author} {\bibfnamefont {J.}~\bibnamefont {{Bobin}}},
  \bibinfo {author} {\bibfnamefont {S.}~\bibnamefont {{Plaszczynski}}},
  \bibinfo {author} {\bibfnamefont {J.-L.}\ \bibnamefont {{Starck}}}, \ and\
  \bibinfo {author} {\bibfnamefont {A.}~\bibnamefont {{Lavabre}}},\ }\href
  {\doibase 10.1051/0004-6361/200912001} {\bibfield  {journal} {\bibinfo
  {journal} {\aap}\ }\textbf {\bibinfo {volume} {519}},\ \bibinfo {eid} {A4}
  (\bibinfo {year} {2010})}\BibitemShut {NoStop}%
\bibitem [{\citenamefont {{Elad}}\ \emph {et~al.}(2005)\citenamefont {{Elad}},
  \citenamefont {{Starck}}, \citenamefont {{Querre}},\ and\ \citenamefont
  {{Donoho}}}]{elad05}%
  \BibitemOpen
  \bibfield  {author} {\bibinfo {author} {\bibfnamefont {M.}~\bibnamefont
  {{Elad}}}, \bibinfo {author} {\bibfnamefont {J.-L.}\ \bibnamefont
  {{Starck}}}, \bibinfo {author} {\bibfnamefont {P.}~\bibnamefont {{Querre}}},
  \ and\ \bibinfo {author} {\bibfnamefont {D.}~\bibnamefont {{Donoho}}},\
  }\href {\doibase 10.1016/j.acha.2005.03.005} {\bibfield  {journal} {\bibinfo
  {journal} {Applied and Computational Harmonic Analysis}\ }\textbf {\bibinfo
  {volume} {19}},\ \bibinfo {pages} {340} (\bibinfo {year} {2005})}\BibitemShut
  {NoStop}%
\bibitem [{\citenamefont {{Hardy}}\ \emph
  {et~al.}(2013{\natexlab{a}})\citenamefont {{Hardy}}, \citenamefont {{Levy}},
  \citenamefont {{M{\'e}tris}}, \citenamefont {{Rodrigues}},\ and\
  \citenamefont {{Touboul}}}]{hardy13a}%
  \BibitemOpen
  \bibfield  {author} {\bibinfo {author} {\bibfnamefont {{\'E}.}~\bibnamefont
  {{Hardy}}}, \bibinfo {author} {\bibfnamefont {A.}~\bibnamefont {{Levy}}},
  \bibinfo {author} {\bibfnamefont {G.}~\bibnamefont {{M{\'e}tris}}}, \bibinfo
  {author} {\bibfnamefont {M.}~\bibnamefont {{Rodrigues}}}, \ and\ \bibinfo
  {author} {\bibfnamefont {P.}~\bibnamefont {{Touboul}}},\ }\href {\doibase
  10.1007/s11214-013-0024-4} {\bibfield  {journal} {\bibinfo  {journal} {\ssr}\
  }\textbf {\bibinfo {volume} {180}},\ \bibinfo {pages} {177} (\bibinfo {year}
  {2013}{\natexlab{a}})}\BibitemShut {NoStop}%
\bibitem [{\citenamefont {{Vitale}}\ \emph {et~al.}(2014)\citenamefont
  {{Vitale}}, \citenamefont {{Congedo}}, \citenamefont {{Dolesi}},
  \citenamefont {{Ferroni}}, \citenamefont {{Hueller}}, \citenamefont
  {{Vetrugno}}, \citenamefont {{Weber}}, \citenamefont {{Audley}},
  \citenamefont {{Danzmann}}, \citenamefont {{Diepholz}}, \citenamefont
  {{Hewitson}}, \citenamefont {{Korsakova}}, \citenamefont {{Ferraioli}},
  \citenamefont {{Gibert}}, \citenamefont {{Karnesis}}, \citenamefont
  {{Nofrarias}}, \citenamefont {{Inchauspe}}, \citenamefont {{Plagnol}},
  \citenamefont {{Jennrich}}, \citenamefont {{McNamara}}, \citenamefont
  {{Armano}}, \citenamefont {{Thorpe}},\ and\ \citenamefont
  {{Wass}}}]{vitale14}%
  \BibitemOpen
  \bibfield  {author} {\bibinfo {author} {\bibfnamefont {S.}~\bibnamefont
  {{Vitale}}}, \bibinfo {author} {\bibfnamefont {G.}~\bibnamefont {{Congedo}}},
  \bibinfo {author} {\bibfnamefont {R.}~\bibnamefont {{Dolesi}}}, \bibinfo
  {author} {\bibfnamefont {V.}~\bibnamefont {{Ferroni}}}, \bibinfo {author}
  {\bibfnamefont {M.}~\bibnamefont {{Hueller}}}, \bibinfo {author}
  {\bibfnamefont {D.}~\bibnamefont {{Vetrugno}}}, \bibinfo {author}
  {\bibfnamefont {W.~J.}\ \bibnamefont {{Weber}}}, \bibinfo {author}
  {\bibfnamefont {H.}~\bibnamefont {{Audley}}}, \bibinfo {author}
  {\bibfnamefont {K.}~\bibnamefont {{Danzmann}}}, \bibinfo {author}
  {\bibfnamefont {I.}~\bibnamefont {{Diepholz}}}, \bibinfo {author}
  {\bibfnamefont {M.}~\bibnamefont {{Hewitson}}}, \bibinfo {author}
  {\bibfnamefont {N.}~\bibnamefont {{Korsakova}}}, \bibinfo {author}
  {\bibfnamefont {L.}~\bibnamefont {{Ferraioli}}}, \bibinfo {author}
  {\bibfnamefont {F.}~\bibnamefont {{Gibert}}}, \bibinfo {author}
  {\bibfnamefont {N.}~\bibnamefont {{Karnesis}}}, \bibinfo {author}
  {\bibfnamefont {M.}~\bibnamefont {{Nofrarias}}}, \bibinfo {author}
  {\bibfnamefont {H.}~\bibnamefont {{Inchauspe}}}, \bibinfo {author}
  {\bibfnamefont {E.}~\bibnamefont {{Plagnol}}}, \bibinfo {author}
  {\bibfnamefont {O.}~\bibnamefont {{Jennrich}}}, \bibinfo {author}
  {\bibfnamefont {P.~W.}\ \bibnamefont {{McNamara}}}, \bibinfo {author}
  {\bibfnamefont {M.}~\bibnamefont {{Armano}}}, \bibinfo {author}
  {\bibfnamefont {J.~I.}\ \bibnamefont {{Thorpe}}}, \ and\ \bibinfo {author}
  {\bibfnamefont {P.}~\bibnamefont {{Wass}}},\ }\href {\doibase
  10.1103/PhysRevD.90.042003} {\bibfield  {journal} {\bibinfo  {journal}
  {\prd}\ }\textbf {\bibinfo {volume} {90}},\ \bibinfo {eid} {042003} (\bibinfo
  {year} {2014})},\ \Eprint {http://arxiv.org/abs/1404.4792} {arXiv:1404.4792
  [gr-qc]} \BibitemShut {NoStop}%
\bibitem [{\citenamefont {{Lomb}}(1976)}]{lomb76}%
  \BibitemOpen
  \bibfield  {author} {\bibinfo {author} {\bibfnamefont {N.~R.}\ \bibnamefont
  {{Lomb}}},\ }\href {\doibase 10.1007/BF00648343} {\bibfield  {journal}
  {\bibinfo  {journal} {Astrophys. and Space Sci.}\ }\textbf {\bibinfo {volume}
  {39}},\ \bibinfo {pages} {447} (\bibinfo {year} {1976})}\BibitemShut
  {NoStop}%
\bibitem [{\citenamefont {{Scargle}}(1982)}]{scargle82}%
  \BibitemOpen
  \bibfield  {author} {\bibinfo {author} {\bibfnamefont {J.~D.}\ \bibnamefont
  {{Scargle}}},\ }\href {\doibase 10.1086/160554} {\bibfield  {journal}
  {\bibinfo  {journal} {\apj}\ }\textbf {\bibinfo {volume} {263}},\ \bibinfo
  {pages} {835} (\bibinfo {year} {1982})}\BibitemShut {NoStop}%
\bibitem [{\citenamefont {{Roberts}}\ \emph {et~al.}(1987)\citenamefont
  {{Roberts}}, \citenamefont {{Lehar}},\ and\ \citenamefont
  {{Dreher}}}]{roberts87}%
  \BibitemOpen
  \bibfield  {author} {\bibinfo {author} {\bibfnamefont {D.~H.}\ \bibnamefont
  {{Roberts}}}, \bibinfo {author} {\bibfnamefont {J.}~\bibnamefont {{Lehar}}},
  \ and\ \bibinfo {author} {\bibfnamefont {J.~W.}\ \bibnamefont {{Dreher}}},\
  }\href {\doibase 10.1086/114383} {\bibfield  {journal} {\bibinfo  {journal}
  {Astron. J.}\ }\textbf {\bibinfo {volume} {93}},\ \bibinfo {pages} {968}
  (\bibinfo {year} {1987})}\BibitemShut {NoStop}%
\bibitem [{\citenamefont {{Foster}}(1995)}]{foster95}%
  \BibitemOpen
  \bibfield  {author} {\bibinfo {author} {\bibfnamefont {G.}~\bibnamefont
  {{Foster}}},\ }\href {\doibase 10.1086/117416} {\bibfield  {journal}
  {\bibinfo  {journal} {Astron. J.}\ }\textbf {\bibinfo {volume} {109}},\
  \bibinfo {pages} {1889} (\bibinfo {year} {1995})}\BibitemShut {NoStop}%
\bibitem [{\citenamefont {{de Waele}}\ and\ \citenamefont
  {{Broersen}}(2000)}]{dewaele00}%
  \BibitemOpen
  \bibfield  {author} {\bibinfo {author} {\bibfnamefont {S.}~\bibnamefont {{de
  Waele}}}\ and\ \bibinfo {author} {\bibfnamefont {P.~M.~T.}\ \bibnamefont
  {{Broersen}}},\ }\href {\doibase 10.1109/78.869039} {\bibfield  {journal}
  {\bibinfo  {journal} {IEEE Transactions on Signal Processing}\ }\textbf
  {\bibinfo {volume} {48}},\ \bibinfo {pages} {2876} (\bibinfo {year}
  {2000})}\BibitemShut {NoStop}%
\bibitem [{Note1()}]{Note1}%
  \BibitemOpen
  \bibinfo {note} {The transpose notation is the common notation in the
  literature of the domain for the direct representation}\BibitemShut {NoStop}%
\bibitem [{\citenamefont {Elad}\ and\ \citenamefont
  {Bruckstein}(2002)}]{elad02}%
  \BibitemOpen
  \bibfield  {author} {\bibinfo {author} {\bibfnamefont {M.}~\bibnamefont
  {Elad}}\ and\ \bibinfo {author} {\bibfnamefont {A.}~\bibnamefont
  {Bruckstein}},\ }\href@noop {} {\bibfield  {journal} {\bibinfo  {journal}
  {IEEE Transactions on Information Theory}\ }\textbf {\bibinfo {volume}
  {48}},\ \bibinfo {pages} {2558} (\bibinfo {year} {2002})}\BibitemShut
  {NoStop}%
\bibitem [{\citenamefont {Donoho}\ and\ \citenamefont {Elad}(2003)}]{donoho03}%
  \BibitemOpen
  \bibfield  {author} {\bibinfo {author} {\bibfnamefont {D.}~\bibnamefont
  {Donoho}}\ and\ \bibinfo {author} {\bibfnamefont {M.}~\bibnamefont {Elad}},\
  }\href@noop {} {\bibfield  {journal} {\bibinfo  {journal} {Proc. Nat. Aca.
  Sci.}\ }\textbf {\bibinfo {volume} {100}},\ \bibinfo {pages} {2197} (\bibinfo
  {year} {2003})}\BibitemShut {NoStop}%
\bibitem [{\citenamefont {{Jullo}}\ \emph {et~al.}(2014)\citenamefont
  {{Jullo}}, \citenamefont {{Pires}}, \citenamefont {{Jauzac}},\ and\
  \citenamefont {{Kneib}}}]{jullo14}%
  \BibitemOpen
  \bibfield  {author} {\bibinfo {author} {\bibfnamefont {E.}~\bibnamefont
  {{Jullo}}}, \bibinfo {author} {\bibfnamefont {S.}~\bibnamefont {{Pires}}},
  \bibinfo {author} {\bibfnamefont {M.}~\bibnamefont {{Jauzac}}}, \ and\
  \bibinfo {author} {\bibfnamefont {J.-P.}\ \bibnamefont {{Kneib}}},\ }\href
  {\doibase 10.1093/mnras/stt2207} {\bibfield  {journal} {\bibinfo  {journal}
  {\mnras}\ }\textbf {\bibinfo {volume} {437}},\ \bibinfo {pages} {3969}
  (\bibinfo {year} {2014})},\ \Eprint {http://arxiv.org/abs/1309.5718}
  {arXiv:1309.5718} \BibitemShut {NoStop}%
\bibitem [{\citenamefont {{Will}}(2014)}]{will14}%
  \BibitemOpen
  \bibfield  {author} {\bibinfo {author} {\bibfnamefont {C.~M.}\ \bibnamefont
  {{Will}}},\ }\href {\doibase 10.12942/lrr-2014-4} {\bibfield  {journal}
  {\bibinfo  {journal} {Living Reviews in Relativity}\ }\textbf {\bibinfo
  {volume} {17}},\ \bibinfo {pages} {4} (\bibinfo {year} {2014})},\ \Eprint
  {http://arxiv.org/abs/1403.7377} {arXiv:1403.7377 [gr-qc]} \BibitemShut
  {NoStop}%
\bibitem [{\citenamefont {{Hardy}}\ \emph
  {et~al.}(2013{\natexlab{b}})\citenamefont {{Hardy}}, \citenamefont {{Levy}},
  \citenamefont {{Rodrigues}}, \citenamefont {{Touboul}},\ and\ \citenamefont
  {{M{\'e}tris}}}]{hardy13b}%
  \BibitemOpen
  \bibfield  {author} {\bibinfo {author} {\bibfnamefont {{\'E}.}~\bibnamefont
  {{Hardy}}}, \bibinfo {author} {\bibfnamefont {A.}~\bibnamefont {{Levy}}},
  \bibinfo {author} {\bibfnamefont {M.}~\bibnamefont {{Rodrigues}}}, \bibinfo
  {author} {\bibfnamefont {P.}~\bibnamefont {{Touboul}}}, \ and\ \bibinfo
  {author} {\bibfnamefont {G.}~\bibnamefont {{M{\'e}tris}}},\ }\href {\doibase
  10.1016/j.asr.2013.07.042} {\bibfield  {journal} {\bibinfo  {journal}
  {Advances in Space Research}\ }\textbf {\bibinfo {volume} {52}},\ \bibinfo
  {pages} {1634} (\bibinfo {year} {2013}{\natexlab{b}})}\BibitemShut {NoStop}%
\bibitem [{\citenamefont {{Salem}}\ and\ \citenamefont
  {{Zygmund}}(1947)}]{salem47}%
  \BibitemOpen
  \bibfield  {author} {\bibinfo {author} {\bibfnamefont {R.}~\bibnamefont
  {{Salem}}}\ and\ \bibinfo {author} {\bibfnamefont {A.}~\bibnamefont
  {{Zygmund}}},\ }\href@noop {} {\bibfield  {journal} {\bibinfo  {journal}
  {Proc. Nat. Acad. Sci.}\ }\textbf {\bibinfo {volume} {33}},\ \bibinfo {pages}
  {333} (\bibinfo {year} {1947})}\BibitemShut {NoStop}%
\end{thebibliography}%
\end{document}